\documentclass[sigconf,10pt]{acmart}

\renewcommand\footnotetextcopyrightpermission[1]{}
\usepackage{graphicx}
\usepackage{amsthm}
\usepackage{tabularx}
\usepackage{multirow}
\usepackage{amsmath} 
\usepackage{pifont} 
\usepackage{upgreek}
\usepackage{wrapfig}
\usepackage{pbox}
\usepackage{bbding}
\usepackage{enumitem}
\usepackage[font=small]{caption}
\usepackage{algorithm}
\usepackage{algpseudocode}

\usepackage{color,soul}
\usepackage[dvipsnames]{xcolor}
\usepackage{xspace}
\usepackage{subfig}
\usepackage{slashbox}
\usepackage[usestackEOL]{stackengine}
\usepackage[normalem]{ulem}
\usepackage{array}
\usepackage{float}
\usepackage{tablefootnote}
\usepackage{tabularray}

\usepackage[utf8]{inputenc}
\usepackage{booktabs, caption, makecell}

\usepackage{threeparttable}

\begin{document}

\setcopyright{acmlicensed}
\copyrightyear{2018}
\acmYear{2018}
\acmDOI{XXXXXXX.XXXXXXX}


\setlength{\abovedisplayskip}{3pt}
\setlength{\belowdisplayskip}{3pt}

\thispagestyle{plain}
\pagestyle{plain}

\definecolor{lightgray}{gray}{0.9}
\definecolor{lightblue}{rgb}{0.9,0.9,1}
\definecolor{LightMagenta}{rgb}{1,0.5,1}
\definecolor{red}{rgb}{1,0,0}

\newcommand\couldremove[1]{{\color{lightgray} #1}}
\newcommand{\remove}[1]{}
\newcommand{\move}[2]{ {\textcolor{Purple}{ \bf --- MOVE #1: --- }} {\textcolor{Orchid}{#2}} }

\newcommand{\hlc}[2][yellow]{ {\sethlcolor{#1} \hl{#2}} }
\newcommand\note[1]{\hlc[SkyBlue]{-- #1 --}} 

\newcommand\mynote[1]{\hlc[yellow]{#1}}
\newcommand\zhihui[1]{\hlc[BurntOrange]{ZG: #1}}
\newcommand\zhecun[1]{\hlc[green]{ZL: #1}}
\newcommand\tingjun[1]{\hlc[yellow]{TC: #1}}

\newcommand{\iu}{{j}}
\newcommand{\eu}{{e}}
\newcommand\usec{$\upmu$s}
\newcommand\usecbf{$\boldsymbol{\upmu}$s}

\newcommand{\myparatight}[1]{\vspace{0.5ex}\noindent\textbf{#1~~}}

\newcommand\name{{\sc BatStation}}
\newcommand\namebf{{\sc\textbf{BatStation}}}

\newcommand{\cmark}{\ding{51}}%
\newcommand{\xmark}{\ding{55}}%
\newcommand{\greencheck}{\color[HTML]{3C8031}{\cmark}}
\newcommand{\redcross}{\color[HTML]{ED1B23}{\xmark}}
\newcommand{\greenno}{\color[HTML]{3C8031}{\textbf{No}}}
\newcommand{\redyes}{\color[HTML]{ED1B23}{\textbf{Yes}}}
\newcommand{\greenlow}{\color[HTML]{3C8031}{\textbf{Low}}}
\newcommand{\redhigh}{\color[HTML]{ED1B23}{\textbf{High}}}

\newcommand{\median}[2]{\underset{#1}{\textsf{median}} \left( #2 \right)}
\newcommand{\hampel}[2]{\underset{#1}{\textsf{Hampel}} \left( #2 \right)}
\newcommand{\round}[1]{\textsf{round}\left( #1 \right)}
\newcommand{\abs}[1]{\left| #1 \right|}

\newcommand{\freqSub}{\Delta {f}}
\newcommand{\fftSize}{N_{\textrm{fft}}}
\newcommand{\sampRate}{f_s}
\newcommand{\symNum}{M}
\newcommand{\symTime}{T_{\textrm{sym}}}
\newcommand{\symIdx}{m}
\newcommand{\subNum}{N}
\newcommand{\subIdx}{n}
\newcommand{\radarNum}{R}
\newcommand{\radarIdx}{r}

\newcommand{\gridRxRaw}{\mathbf{Y}_{\textrm{raw}}}
\newcommand{\gridRxRes}{\mathbf{Y}_{\textrm{res}}}

\newcommand{\gridTxComm}{\mathbf{X}_{\textrm{5G}}}
\newcommand{\gridTxCommEst}{\widehat{\mathbf{X}}_{\textrm{5G}}}
\newcommand{\gridRxComm}{\mathbf{Y}_{\textrm{5G}}}
\newcommand{\gridRxCommEst}{\widehat{\mathbf{Y}}_{\textrm{5G}}}
\newcommand{\gridRxRadar}{\mathbf{Y}_{\textrm{radar}}}
\newcommand{\gridRxNoise}{\mathbf{Y}_{\textrm{noise}}}

\newcommand{\specTxComm}{\mathbf{x}_{\textrm{5G}}}
\newcommand{\specTxCommEst}{\widehat{\mathbf{x}}_{\textrm{5G}}}
\newcommand{\specTxCommData}{\mathbf{x}_{\textrm{5G}}^{\textrm{data}}}
\newcommand{\specTxCommDataEst}{\widehat{\mathbf{x}}_{\textrm{5G}}^{\textrm{data}}}
\newcommand{\specTxCommDMRS}{\mathbf{x}_{\textrm{5G}}^{\textrm{DMRS}}}
\newcommand{\specTxCommDMRSEst}{\widehat{\mathbf{x}}_{\textrm{5G}}^{\textrm{DMRS}}}
\newcommand{\specTxRawData}{\mathbf{x}_{\textrm{raw}}^{\textrm{data}}}
\newcommand{\specTxRawDataEst}{\widehat{\mathbf{x}}_{\textrm{raw}}^{\textrm{data}}}
\newcommand{\specRxRaw}{\mathbf{y}_{\textrm{raw}}}
\newcommand{\specRxRawData}{\mathbf{y}_{\textrm{raw}}^{\textrm{data}}}
\newcommand{\specRxRawDataEst}{\widehat{\mathbf{y}}_{\textrm{raw}}^{\textrm{data}}}
\newcommand{\specRxRawDMRS}{\mathbf{y}_{\textrm{raw}}^{\textrm{DMRS}}}
\newcommand{\specRxRawDMRSEst}{\widehat{\mathbf{y}}_{\textrm{raw}}^{\textrm{DMRS}}}
\newcommand{\specRxCommData}{\mathbf{y}_{\textrm{5G}}^{\textrm{data}}}
\newcommand{\dmrsWithm}{\mathbf{x}_{\textrm{5G}}^{\textrm{DMRS}}}
\newcommand{\specRxCommDataEst}{\widehat{\mathbf{y}}_{\textrm{5G}}^{\textrm{data}}}
\newcommand{\specRxCommDMRS}{\mathbf{y}_{\textrm{5G}}^{\textrm{DMRS}}}
\newcommand{\specRxCommDMRSEst}{\widehat{\mathbf{y}}_{\textrm{5G}}^{\textrm{DMRS}}}
\newcommand{\specRxComm}{\mathbf{y}_{\textrm{5G}}}
\newcommand{\specRxCommEst}{\widehat{\mathbf{y}}_{\textrm{5G}}}
\newcommand{\specRxRes}{\mathbf{y}_{\textrm{res}}}
\newcommand{\specCSIRaw}{\mathbf{h}_{\textrm{raw}}}
\newcommand{\specCSIRawEst}{\widehat{\mathbf{h}}_{\textrm{raw}}}
\newcommand{\specCSIComm}{\mathbf{h}_{\textrm{5G}}}
\newcommand{\specCSICommEst}{\widehat{\mathbf{h}}_{\textrm{5G}}}

\newcommand{\gridRxOpt}{Y^{\star}}
\newcommand{\radarIdxOpt}{r^{\star}}
\newcommand{\subIdxOpt}{n^{\star}}
\newcommand{\symIdxOpt}{m^{\star}}
\newcommand{\outputDetectThres}{Y_{\textrm{th}}}
\newcommand{\outputDetect}{D}
\newcommand{\outputDetectEst}{\widehat{D}}
\newcommand{\outputClass}{R}
\newcommand{\outputClassEst}{\widehat{R}}
\newcommand{\outputLocalFreq}{L_{\textrm{freq}}}
\newcommand{\outputLocalFreqEst}{\widehat{L}_{\textrm{freq}}}
\newcommand{\outputLocalTime}{L_{\textrm{time}}}
\newcommand{\outputLocalTimeEst}{\widehat{L}_{\textrm{time}}}
\newcommand{\prob}[1]{\textsf{Pr}\left( #1 \right)}

\newcommand{\refftNum}{\alpha}
\newcommand{\gridRxFFTSet}{\mathcal{Y}_{\textrm{fft}}}
\newcommand{\gridRxFFT}{\mathbf{Y}_{\textrm{fft}}}
\newcommand{\poolSize}{P}
\newcommand{\gridRxPoolSet}{\mathcal{Y}_{\textrm{pool}}}
\newcommand{\gridRxPool}{\mathbf{Y}_{\textrm{pool}}}

\newcommand{\templateSet}{\mathcal{T}}
\newcommand{\template}{\mathbf{T}}
\newcommand{\layerNum}{C}
\newcommand{\layerIdx}{c}

\newcommand{\gridRxOutSet}{\mathcal{Y}_{\textrm{out}}}
\newcommand{\gridRxOut}{\mathbf{Y}_{\textrm{out}}}

\newcommand{\dataIdx}{i}
\newcommand{\pattern}{\mathbf{P}}

\title{{\namebf}: Toward In-Situ Radar Sensing on 5G Base Stations with Zero-Shot Template Generation}

\author{Zhihui Gao, Zhecun Liu, Tingjun Chen}
\affiliation{%
  \institution{\vspace{0.5ex} Department of Electrical and Computer Engineering, Duke University\vspace{0.5ex}}
  \country{}
}
\email{{zhihui.gao, zhecun.liu, tingjun.chen}@duke.edu}

\begin{abstract}
The coexistence between incumbent radar signals and commercial 5G signals necessitates a versatile and ubiquitous radar sensing for efficient and adaptive spectrum sharing.
In this context, leveraging the densely deployed 5G base stations (BS) for radar sensing is particularly promising, offering both wide coverage and immediate feedback to 5G scheduling.
However, the targeting radar signals are superimposed with concurrent 5G uplink transmissions received by the BS, and practical deployment also demands a lightweight, portable radar sensing model.
This paper presents {\name}, a lightweight, in-situ radar sensing framework seamlessly integrated into 5G BSs.
{\name} leverages uplink resource grids to extract radar signals through three key components:
(\emph{i}) radar signal separation to cancel concurrent 5G transmissions and reveal the radar signals,
(\emph{ii}) resource grid reshaping to align time-frequency resolution with radar pulse characteristics, and
(\emph{iii}) zero-shot template correlation based on a portable model trained purely on synthetic data that supports detection, classification, and localization of radar pulses without fine-tuning using experimental data.
We implement {\name} on a software-defined radio (SDR) testbed and evaluate its performance with real 5G traffic in the CBRS band. 
Results show robust performance across diverse radar types, achieving detection probabilities of {97.02\%} (PUCCH) and {79.23\%} (PUSCH), classification accuracy up to {97.00\%}, and median localization errors of {2.68--6.20}\thinspace{MHz} (frequency) and {24.6--32.4}\thinspace{\usec} (time). 
Notably, {\name} achieves this performance with a runtime latency of only {0.11/0.94}\thinspace{ms} on GPU/CPU, meeting the real-time requirement of 5G networks.
\end{abstract}

\maketitle

\section{Introduction}
\label{sec: introduction}

The cellular networks in the 5G/beyond-5G era maneuver large bandwidth to meet the increasingly large data rate demand, which expands to the spectrum co-existing with other incumbent services~\cite{Parastar2023, Khan2024, Zhang2023}.
For example, announced by the federal communications commission (FCC) recently, the citizens broadband radio service (CBRS) band located at {3.55--3.70}\thinspace{GHz} is released for commercial 5G access, but at a lower priority than the incumbent access (e.g., satellite earth stations, and the U.S. Navy). 
Such a hierarchical access means that the 5G communications are supposed to detect the incumbent occupancy, and either avoid collision or minimize the interference.
Among the incumbent occupancies, the naval radar managed by the department of defense (DoD) commonly occurs, whose radar signals are challenging to detect due to their deployment on mobile radar transceivers, as well as the low signal power.

To accomplish this, a tremendous amount of effort~\cite{urkowitz1967energy, Bhattacharya2020, sarkar2021deepradar, sarkar2024radyololet, doke2024radview, ghosh2024sparc, reus2023senseoran} has been devoted to developing dedicated radar sensors to sense the radar signals.
Fundamentally, the radar signal's existence can be determined by thresholding the received energy either directly in the time domain signals or on the spectrogram after the short-time Fourier transform (STFT).
Recently, advanced machine learning (ML) driven methods~\cite{sarkar2021deepradar, sarkar2024radyololet, doke2024radview, ghosh2024sparc}, mainly derived from YOLO~\cite{redmon2016you}, extract detailed radar information from the spectrogram, such as the radar type, center frequency, and starting time.

\begin{figure}[!t]
    \centering
    \includegraphics[width=1.0\columnwidth]{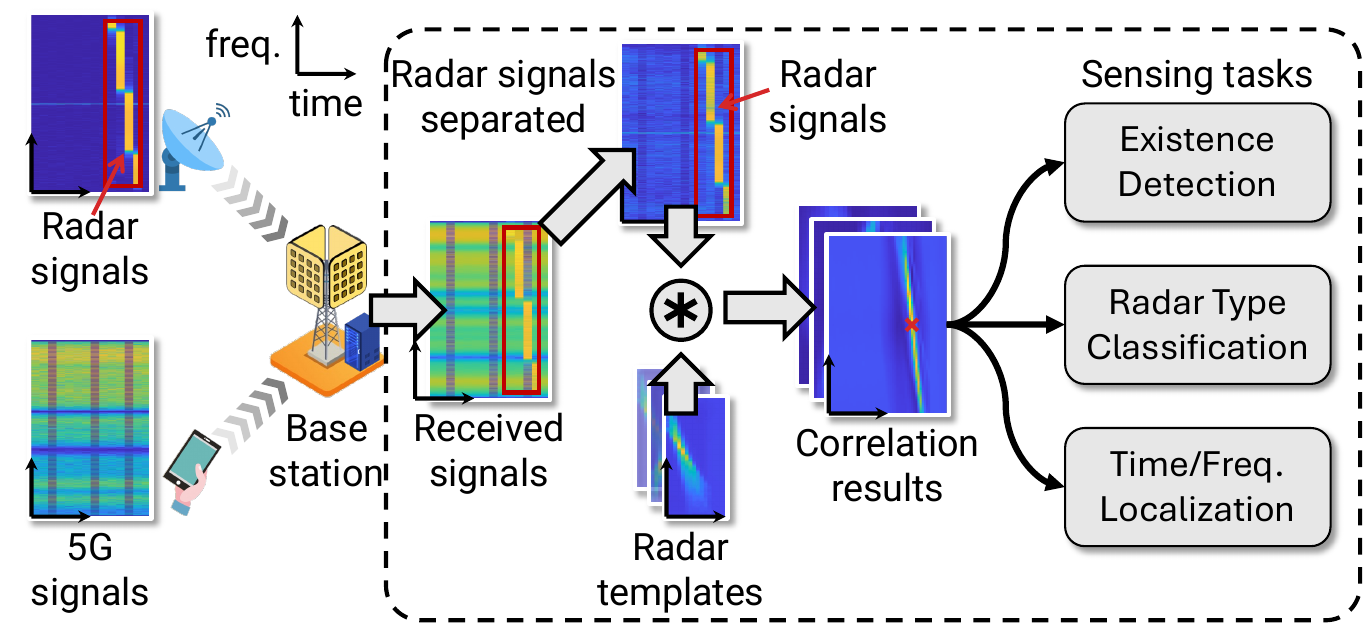}
    \vspace{-8mm}
    \caption{In shared spectrum bands such as CBRS, the radar signals and 5G signals may interfere, resulting in superimposed spectrograms received by the BS. 
    {\namebf} separates the radar signals, and correlates them with zero-shot generated radar templates, whose results drive sensing tasks of (\emph{i}) radar existence detection, (\emph{ii}) radar type classification, and (\emph{iii}) starting time and center frequency localization.}
    \vspace{-5mm}
    \label{fig: system-introduction}
\end{figure}

Beyond using dedicated radar sensors, the rollout of 5G networks with densely deployed base stations (BSs) provides an alternative solution for radar sensing in the same/adjacent frequency bands.
During uplink transmissions, BSs receive not only 5G signals from the users but also potential radar signals, thus serving as opportunistic radar sensors.
Embedding such \emph{in-situ} radar sensing capability directly into 5G BSs not only eliminates the need for dedicated radar sensors, but also facilitates instant spectrum-aware 5G scheduling~\cite{baldesi2022charm, ghosh2024sparc}.

\begin{table*}[!t]
\begin{threeparttable}
\caption{Comparison of existing radar sensing systems with the proposed systems, {\namebf}.}
\label{tab:system-comparison}
\footnotesize
\vspace{-3mm}
\begin{tabular}{|c|c|c|c|c|c|c|c|}
    \hline
    & \multicolumn{3}{c|}{\textbf{Functionality}} & \multicolumn{3}{c|}{\textbf{Capability}} & {\textbf{Evaluation}} \\
    \cline{2-8}
    & \Centerstack{Existence \\ Detection}
    & \Centerstack{Radar Type \\ Classification}
    & \Centerstack{Time/Frequency \\ Localization}
    & \Centerstack{5G Protocol \\ Compatibility}
    & \Centerstack{Runtime \\ Efficiency}
    & \Centerstack{Zero-Shot \\ Generation}
    & \Centerstack{Experiment \\ Conducted}\\
    \hline
    Energy Detection~\cite{urkowitz1967energy} & {\greencheck} & {\redcross} & {\greencheck} & {\redcross} & {\greencheck} & {\greencheck} & {\greencheck} \\
    SPARC~\cite{ghosh2024sparc} & {\greencheck} & {\redcross} & {\greencheck} & {\greencheck} & {\redcross} & {\redcross} & {\greencheck} \\
    DeepRadar~\cite{sarkar2021deepradar} & {\greencheck} & {\greencheck} & {\greencheck} & {\redcross} & {\redcross} & {\redcross} & {\greencheck} \\
    RadVIEW~\cite{doke2024radview} & {\greencheck} & {\redcross} & {\greencheck} & {\redcross} & {\redcross} & {\redcross} & {\redcross} \\
    RadYOLO~\cite{sarkar2024radyololet} & {\greencheck} & {\greencheck} & {\greencheck} & {\redcross}& {\greencheck} & {\redcross} & {\redcross} \\
    SenseORAN~\cite{reus2023senseoran} & {\greencheck} & {\greencheck} & {\greencheck} & {\greencheck} & {\redcross} & {\redcross} & {\greencheck} \\
    {\namebf} & {\greencheck} & {\greencheck} & {\greencheck} & {\greencheck} & {\greencheck} & {\greencheck} & {\greencheck} \\
    \hline
\end{tabular}
\end{threeparttable}
\vspace{-3mm}
\normalsize
\end{table*}

Compared to dedicated radar sensors, however, enabling radar sensing at 5G BSs poses two key challenges.
First, \emph{radar signals suffer from (potentially severe) interference caused by coexisting 5G uplink transmissions}.
In shared spectrum bands, 5G signals may overlap with radar signals in both frequency and time, especially in 5G's physical uplink shared channel (PUSCH) under heavy traffic conditions.
Moreover, the 5G uplink signal--acting as interference to radar signals--may lead to a low interference-plus-noise ratio (SINR) for radar sensing.
Fortunately, we notice that as long as the uplink data can be correctly decoded, the corresponding 5G signals can be reverse-engineered and canceled, thereby revealing the radar signals.
Such radar signal separation can significantly enhance the SINR available for radar sensing.

Second, \emph{radar sensing must be performed in real-time to capture fast-varying signals}, which is critical to avoid delaying feedback to the 5G scheduling process.
However, many existing radar sensing approaches~\cite{ghosh2024sparc, sarkar2021deepradar, doke2024radview, reus2023senseoran}, particularly those based on ML, often suffer from high latency due to the combination of the substantial throughput of the received signals and the complicated ML model architecture with heavy multiply-accumulates (MAC) operations involved during ML inference.
In addition, training such ML models typically requires extensive experimental data, which is often costly to obtain.
To address this issue, we notice that the appearance of the radar signals on the 5G resource grids (i.e., the spectrogram in 5G by a specific STFT rule) follows certain patterns for each radar type. 
Thereby, one can derive a set of templates corresponding to different radar types that capture their unique ``patterns''.
When correlating the templates with the resource grids, a correlation peak appears.
Compared to those sophisticated ML models, this template correlation process is equivalent to a single 2D convolutional layer without nonlinear activation functions involved.

In this paper, we propose {\name}\footnote{Bats do a very job at detecting chirps for localization and navigation.}, a radar sensing system on 5G BSs in the co-existing band, which functions for radar existence detection, radar type classification (from which the radar parameters, such as the duration and bandwidth, can be inferred), and radar center frequency/starting time localization, as illustrated in Fig.~\ref{fig: system-introduction}.
By leveraging the radar sensing information, the BS can identify the spectrum occupancy of radar signals, and subsequently avoid scheduling 5G signals that may cause interference in the future. 
{\name} consists of three modules: (\emph{i}) \emph{Radar Signal Separation}, which separates and reveals the radar signals on the 5G resource grids from concurrent uplink transmissions;
(\emph{ii}) \emph{Resource Grid Reshaping}, which reshapes and compresses the time and frequency resolutions of the resource grid to incorporate the target radar bandwidth/duration; and
(\emph{iii}) \emph{Radar Template Correlation}, which detects, classifies, and localizes the radar signals via correlation with a set of radar templates optimized and trained on purely synthetic data (i.e., zero-shot template generation).
A comparison of {\name} to existing radar sensing systems is summarized in Table~\ref{tab:system-comparison}.

We implement and experimentally evaluate {\name} on a software-defined radio (SDR) platform operating in the CBRS band at {3.55--3.70}\thinspace{GHz}, where 5G traffic between USRP N321 (BS) and commodity smartphones (users) is generated by Open5GS~\cite{open5gs} and srsRAN~\cite{srsgithub}, and the radar signals are generated by an auxiliary USRP X310 SDR.
Extensive experiments demonstrate that, under PUCCH with concurrent 5G uplink transmissions, {\name} can detect all five radar types radar signals with a mean detection probability of {97.02\%}, which decreases to {79.23\%} under PUSCH with heavy 5G traffic with an interference-to-noise ratio (INR) of {24.3--38.4}\thinspace{dB}.
Meanwhile, {\name} achieves radar type classification accuracies of {97.00\%} and {95.30\%} for PUCCH and PUSCH, respectively.
Similarly, for PUCCH/PUSCH, the median error of the center frequency localization is {2.68/6.20}\thinspace{MHz} (i.e., {2.68/6.20\%} of a {100}\thinspace{MHz} channel) and the median error of the starting time localization is {24.6/32.4}\thinspace{\usec} (i.e., {4.91/6.48\%} of a {0.5}\thinspace{ms} uplink slot).
Furthermore, with only {4,560} parameters and {40}\thinspace{M} MAC operations per inference, the latency of {\name} is {0.11/0.90}\thinspace{ms} on CPU/GPU, supporting the real-time deployment on the 5G BS.
Beyond the CBRS band, {\name} can also be readily adapted to other spectrum bands and radar types.

To summarize, the main contributions of this paper are:
\begin{itemize}[leftmargin=*, topsep=2pt, itemsep=1pt]
    \item
    \emph{In-situ radar sensing on 5G BSs}: {\name} enables radar sensing directly on 5G base stations during uplink transmissions, ensuring compatibility with 3GPP standards;
    \item
    \emph{Lightweight and portable template correlation model}: {\name} introduces a zero-shot, template-based correlation method that is computationally lightweight and agnostic to radio hardware;
    \item
    \emph{Extensive experimental validation}: {\name} is evaluated using an SDR platform with real 5G traffic, demonstrating robust detection, classification, and localization across multiple radar types with varying parameters.
\end{itemize}

\section{Related Work}
\label{sec:related}

\myparatight{Spectrum sensing in co-existence band.}
Recent efforts have been devoted to examining the signal occupancy on the spectrum across time and frequency.
The common technology to sense the general signals includes sweeping the frequency of a specialized oscillator~\cite{guddeti2019sweepsense, subbaraman2023crescendo}, and unfolding the frequency aliasing by compressive sensing~\cite{Song2023a} or deep learning~\cite{peng2024sums}.
Specifically for Wi-Fi signals, SWIRLS~\cite{gao2023swirls} and SigChord~\cite{peng2025sigchord} (partially) decode the Wi-Fi packets for finer-grained information, while NG-scope~\cite{xie2022ng} is proposed to monitor the 5G signals' spectrum usage.
This spectrum sensing knowledge facilitates the spectrum sharing in the cellular networks~\cite{CalvoPalomino2020a, uvaydov2021deepsense, ko2024edgeric, Fiore2023, Zumegen2024, Ichkov2025}, especially in the co-existence band~\cite{Baig2018, Xu2022, Hammouda2019}.
In particular, CHARM~\cite{baldesi2022charm} and SPARC~\cite{ghosh2024sparc} integrate it into the scheduling of LTE and 5G networks; there are existing works~\cite{zeng2023adaptive, ko2024edgeric, santhi2025interfo} that further accelerate the process to meet the real-time requirement.

\myparatight{Radar sensing systems.}
Traditionally, the radar sensing can be categorized into energy detection~\cite{urkowitz1967energy, bell2023searchlight}, and matched filters~\cite{caromi2018detection}.
Recently, the development of ML, especially the YOLO series~\cite{redmon2016you}, has largely enhanced the detection capability and further allows fine-grained radar parameter estimation, such as bandwidth, duration~\cite{sarkar2021deepradar, doke2024radview, sarkar2024radyololet, luo2022wise}.
Close to {\name}, SenseORAN~\cite{reus2023senseoran} integrates radar sensing to 5G BS in the assistance of ML. However, these ML systems require an experimental dataset that aligns with the real-world deployment, and are usually computationally heavy.

\myparatight{Novel applications on cellular networks}
The wide deployment of the 5G BSs attracts a variety of novel applications while maintaining the communication functionality, such as the localization of either user devices~\cite{zhu2023experience, oh2024enabling, garg2024litefoot, Lizarribar2024, Schott2024a}, human~\cite{Shastri2024}, or mobile vehicles~\cite{Li2022, LimaniFazliu2022}.
In addition, the emerging mmWave spectrum together with the large-scale antenna arrays on the BS incorporates beamforming, further enhancing the sensing capability, including imaging~\cite{guan20213}, augmented reality~\cite{Ghoshal2023}, or long-range sensing~\cite{Mohan2023, Guan2022}.
\section{Preliminaries}
\label{sec:preliminaries}

\begin{figure*}[!h]
    \begin{minipage}{0.75\columnwidth}
        \centering
        \vspace{10mm}
        \includegraphics[width=0.98\columnwidth]{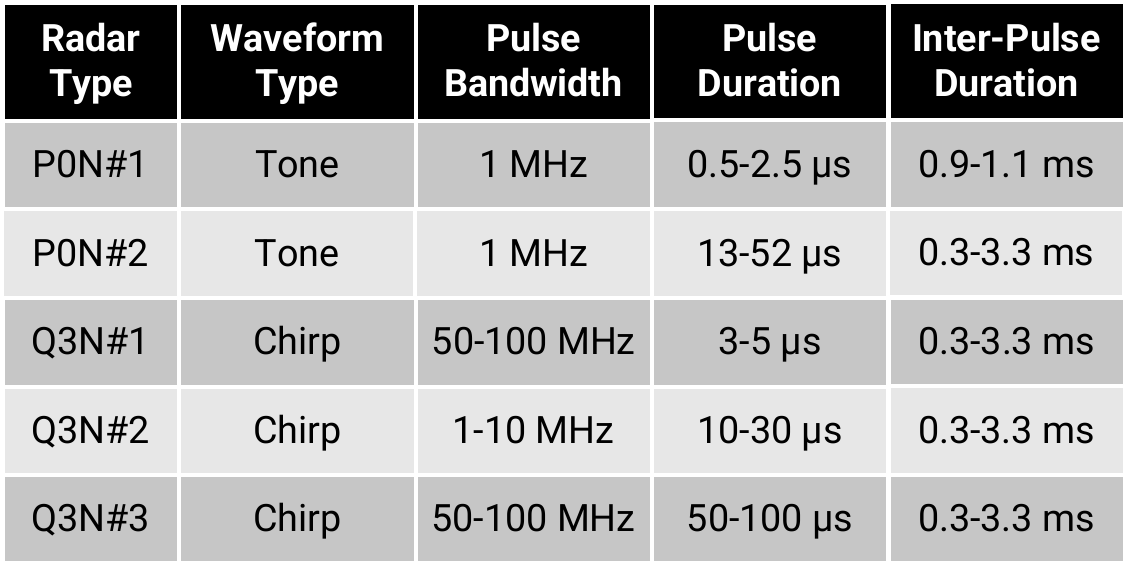}
        \vspace{8mm}
        \vspace{-3mm}
        \captionof{table}{The parameters of the radar types co-existing on the CBRS band.}
        \vspace{-3mm}
        \label{fig:radar-characterstics}
    \end{minipage}
    \hfill
    \begin{minipage}{1.30\columnwidth}
        \centering
        \includegraphics[width=0.98\columnwidth]{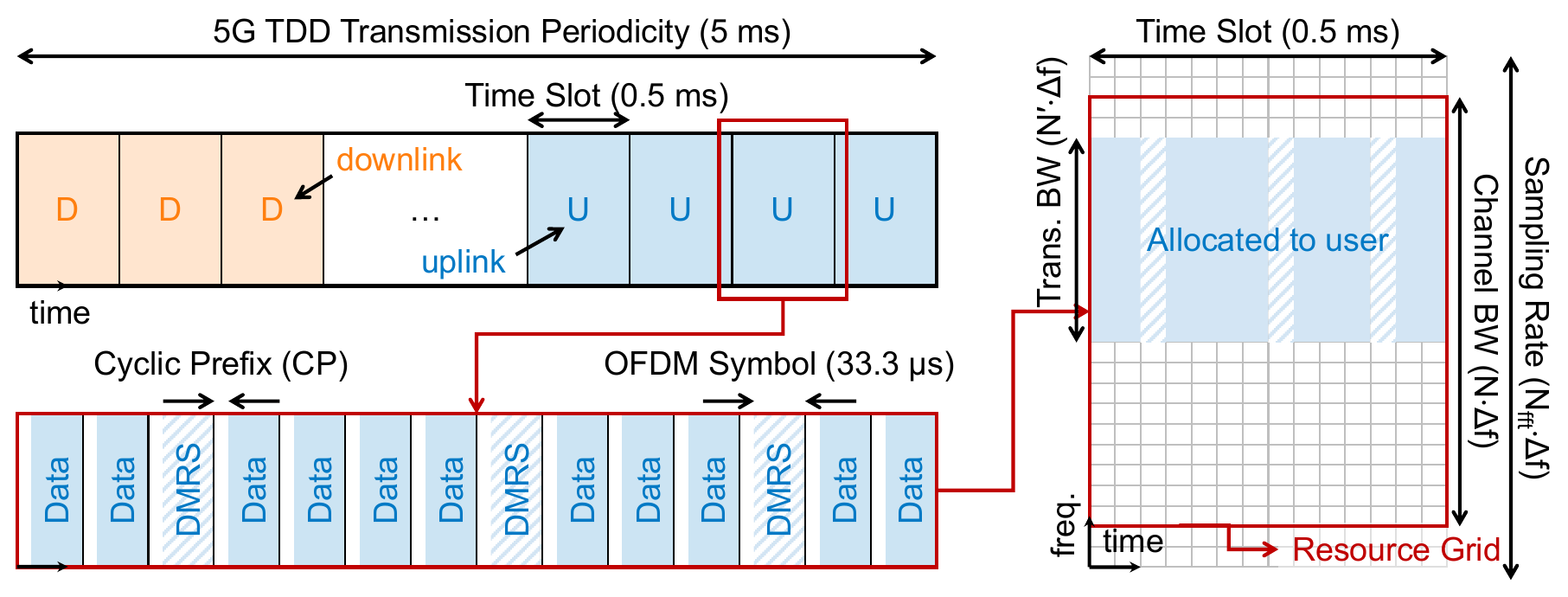}
        \vspace{-3mm}
        \captionof{figure}{The 5G time/frequency structure: a resource grid is dimensioned by the subcarrier number in frequency and the OFDM symbol number in time.}
        \vspace{-3mm}
        \label{fig:resource-grid}
    \end{minipage}
    \vspace{-3mm}
\end{figure*}

\subsection{Radar Signals in the CBRS Band}
\label{sec:coexisted-radar}

In this paper, we focus on the CBRS band (3.55--3.7\thinspace{GHz})~\cite{caromi2019rf, sanders2017procedures} with five types of radar signals.
Note that {\name} is a general radar sensing system that can be easily extended to other radar types and frequency bands.

\myparatight{Radar types.}
The coexisting radar signals appear in the form of \emph{burst}, which is composed of multiple repeated \emph{pulses} with the same time intervals; the radar pulses within a given burst have the identical waveform.
As described in Fig.~\ref{fig:radar-characterstics}, there are five radar types with their own parameter ranges~\cite{caromi2019rf, sanders2017procedures}, namely, P0N\#1, P0N\#2, and Q3N\#1, Q3N\#2, and Q3N\#3, where the first two types are backboned by the tone waveforms, and the last three types are by chirp waveforms.
Specifically, these five radar types can be distinguished by their bandwidth and/or duration ranges.
Note that the three chirp-based radar types (Q3N\#1, Q3N\#2, and Q3N\#3) can either sweep from low frequency to high (up-chirp) or in the reverse direction (down-chirp).

\myparatight{Radar signal SINR.}
The signal-to-noise ratio (SNR) or signal-to-interference-plus-noise ratio (SINR) of the radar signals is defined as the power spectral density (PSD) ratio within the radar pulses' occupied bandwidth and duration.
This SNR/SINR definition for radar signals is generally higher than that in the 5G networks, whose noise power is aggregated over all the receiving bandwidth and time.
In {\name}, the 5G signals serve as the interference when estimating the SINR of the radar signals.

\subsection{5G Physical Layer (PHY)}

A 5G network is based on the orthogonal frequency-division multiplexing (OFDM), which is featured by the fast-Fourier transform (FFT) size of $\fftSize$ (e.g., {4,096}) and subcarrier spacing $\freqSub$. Without loss of generality, we take the 5G numerology 1 (NU1) in this paper, where $\freqSub$={30}\thinspace{kHz}.

\myparatight{5G time-frequency resource grid.}
Today's 5G networks follow time-division duplex (TDD), which switches the downlink and uplink transmissions in the time domain while sharing the same spectrum~\cite{sharetechnote, Ghoshal2022, Zeng2023}.
As shown in Fig.~\ref{fig:resource-grid}, the smallest period of downlink-uplink switching, \emph{transmission periodicity}, lasts for {5}\thinspace{ms}, which starts with downlink slots and ends with uplink slots of {0.5}\thinspace{ms} each.
The uplink slots can be categorized by functionality into \emph{physical uplink control channel (PUCCH)} to convey control information with light traffic, and \emph{physical uplink shared channel (PUSCH)} to convey the body data information.
Each uplink slot is composed of $\symNum={14}$ OFDM symbols in time, and each OFDM symbol is composed of $\fftSize$ subcarriers spaced at $\freqSub={30}\thinspace\textrm{kHz}$ in the frequency domain. By conducting the $\fftSize$-point IFFT, it becomes $\fftSize$ complex-valued I/Q samples lasting for $\symTime=1/\freqSub={33.3}\thinspace\textrm{\usec}$ (excluding the cyclic prefix) in the time domain.
Among all $\fftSize$ subcarriers, only the middle $\subNum$ subcarriers (corresponding to a channel bandwidth of $\subNum \cdot \freqSub$) can be allocated for communication in demand.
Hence, the 5G resource grid of a time slot can be formulated as a complex-valued matrix, $\gridTxComm \in \mathbb{C}^{\subNum \times \symNum}$; specifically, for one column in $\gridTxComm$, denoted as $\specTxComm \in \mathbb{C}^{\subNum}$, is the frequency-domain representation of an OFDM symbol.
Within a resource grid, a user is allocated a subset of subcarriers in demand, whose total bandwidth is equal to the \emph{transmission bandwidth}.
In practice, the number of allocated subcarriers depends on the user's communication demands, which are usually larger in PUSCH than in PUCCH.

\myparatight{DMRS symbols and data demodulation.}
Within an uplink slot, there exist two types of OFDM symbols: \emph{data symbols}, where the uplink data is modulated, and \emph{demodulation reference signal (DMRS) symbols}, where per-subcarrier channel state information (CSI) can be estimated to demodulate the data symbols.
As a concrete example, Fig.~\ref{fig:resource-grid} depicts the so-called Type-1 DMRS format~\cite{sharetechnote}, where the three DMRS symbols appear at $3^{rd}$, $8^{th}$, and $12^{th}$ symbols, and the rest are the data symbols.
Specifically, let $\specTxCommData \in \mathbb{C}^{\subNum}$, $\dmrsWithm \in \mathbb{C}^{\subNum}$ denote the transmitted data or DMRS symbols from the user, and $\specRxCommData \in \mathbb{C}^{\subNum}$, $\specRxCommDMRS \in \mathbb{C}^{\subNum}$ the respective received symbols at the BS. 
By comparing it to the received DMRS symbol $\specRxCommDMRS$, the per-subcarrier CSI, denoted as $\specCSICommEst \in \mathbb{C}^{\subNum}$, can be estimated by $\specCSICommEst = \specRxCommDMRS \oslash \specTxCommDMRS$, where $\oslash$ is the element-wise division.
Assuming the stable CSI over time, the transmitted data symbols, denoted as $\specTxCommDataEst$, can be calibrated by the BS as $\specTxCommDataEst = \specRxCommData \oslash \specCSICommEst$.
\section{Problem Formulation}

{\name} provides three radar sensing functionalities: radar existence detection, radar type classification, and center frequency/starting time localization.
Unlike prior works, {\name} achieves these functionalities \emph{in-situ} at the BS by analyzing individual radar pulses during uplink slots.

\myparatight{BS's received 5G resource grid.}
The 5G BS receives the time domain I/Q samples, and converts them to the complex-valued resource grids using the $\fftSize$-point FFT.
We denote the resource grids for one uplink slot as $\gridRxRaw \in \mathbb{C}^{\subNum \times \symNum}$, serving as the input of {\name}.

\myparatight{Radar existence detection.}
The radar detection problem can be formulated as a binary classification task: determining whether a radar pulse is present in the resource grid $\gridRxRaw$, regardless of the radar types or parameters.
We denote the detection output of {\name} by $\outputDetect \in \{ 0, 1 \}$, where $\outputDetect=1$ indicates the presence of a radar pulse, and $0$ otherwise.

\myparatight{Radar type classification.}
{\name} also supports radar type classification, i.e., identifying the specific radar type listed in Table~\ref{fig:radar-characterstics}.
This task is formulated as a five-class classification problem, where the output label $\outputClass \in \{ 1, 2, 3, 4, 5 \}$ corresponds to radar types P0N\#1, P0N\#2, Q3N\#1, Q3N\#2, and Q3N\#3, respectively.

\myparatight{Center frequency and starting time localization.}
Furthermore, {\name} is capable of localizing the center frequency (or center subcarrier index) and the starting time (or the starting OFDM symbol index) of the radar pulse, which are formulated as two independent regression tasks.
Specifically, we denote the radar signal center frequency as $\outputLocalFreq \in \mathbb{R}$, and starting time as $\outputLocalTime \in \mathbb{R}$.
\section{System Design}
\label{sec:system-design}

\begin{figure*}[!t]
    \centering
    \includegraphics[width=0.9\textwidth]{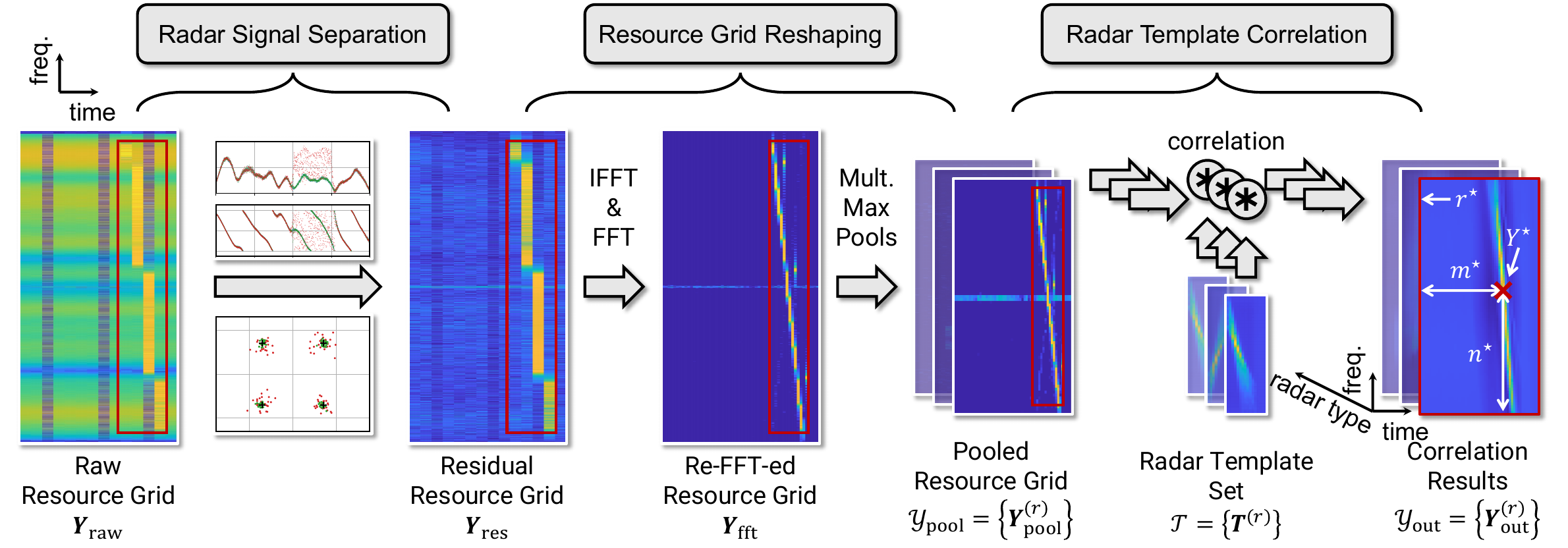}
    \vspace{-3mm}
    \caption{{\namebf}'s workflow: first, the radar signals are separated from the raw resource grid $\gridRxRaw$; then, the separated resource grid $\gridRxRes$ is reshaped by re-conducting FFT and different levels of max pooling along the frequency dimension; finally, the set of pooled resource grids $\gridRxPoolSet$ is respectively correlated with a set of radar templates $\templateSet$, and the radar sensing results can be obtained by finding the maximum values $\gridRxOpt$ located at $(\radarIdxOpt, \subIdxOpt, \symIdxOpt)$ on the correlation results $\gridRxOutSet$.}
    \vspace{-3mm}
    \label{fig:system-overview}
\end{figure*}

{\name} consists of three main modules (Fig.~\ref{fig:system-overview}):
(\emph{i}) \emph{Radar Signal Separation}, which separates and reveals the radar signals on the raw input resource grid from concurrent 5G transmissions;
(\emph{ii}) \emph{Resource Grid Reshaping}, which reshapes and compresses the time and frequency resolution of the resource grid to incorporate the targeting radar bandwidth/duration and reduce computation cost, and 
(\emph{iii}) \emph{Radar Template Correlation with Zero-Shot Template Generation}, which detects, classifies, and localizes the radar signals via correlation with a set of radar templates optimized and trained on purely synthetic data (i.e., zero-shot template generation).

\subsection{Radar Signal Separation}
\label{ssec: design-separation}

\begin{figure}[!t]
    \centering
    \includegraphics[width=0.95\columnwidth]{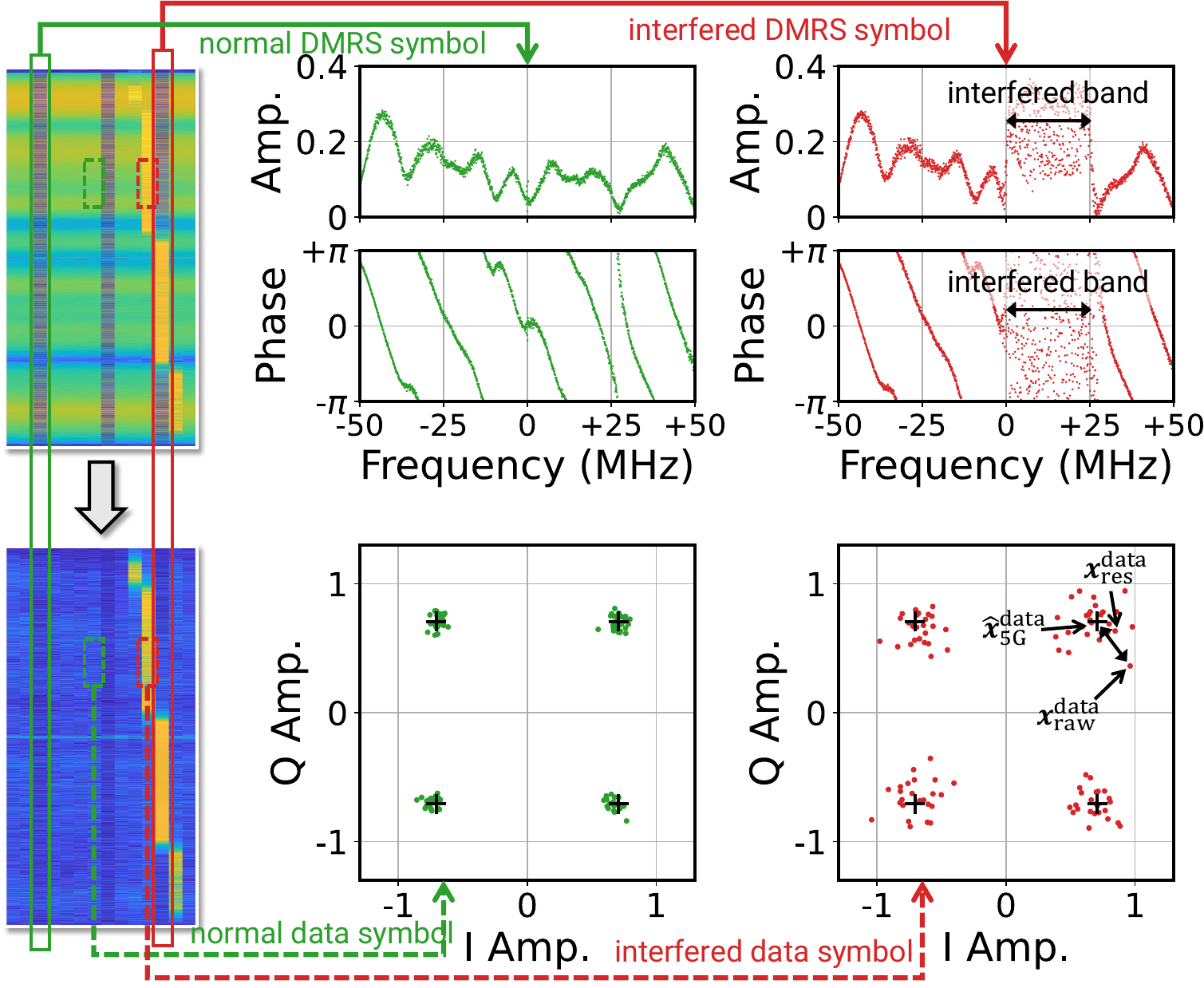}
    \vspace{-3mm}
    \caption{(\emph{Top}) The per-subcarrier CSI estimated using normal and interfered DMRS symbols.
    (\emph{Bottom}) The constellations of normal and interfered data symbols.}
    \vspace{-5mm}
    \label{fig: radar-separation}
\end{figure}

This \emph{Radar Signal Separation} module subtracts the 5G signals, $\gridRxComm$, from the raw resource grid $\gridRxRaw$ received by the BS, yielding the residual resource grid, denoted as $\gridRxRes$, with the radar signals $\gridRxRadar$ and noises $\gridRxNoise$ only, as shown in Fig.~\ref{fig:system-overview}.

\myparatight{Exploring radar signal properties and 5G transmission schedule.}
According to Table~\ref{fig:radar-characterstics}, the inter-pulse interval of the five radar types is {0.33--3.33}\thinspace{ms}, which is much longer than the OFDM symbol duration of {33.3}\thinspace{\usec} in NU1.
Therefore, only a small portion of the DMRS symbols are interfered by the radar pulse, while the rest can still provide correct CSI estimation.
Given the correct CSI, the data symbols can be (mostly) correctly decoded as long as the interference from the radar signals is not strong.
In this way, the resource grid corresponding to the 5G uplink transmissions $\gridRxComm$--including both DMRS and data symbols--can be reverse-engineered and then subtracted from $\gridRxRaw$ to reveal the radar signal.

\myparatight{CSI estimation based on interfered DMRS symbols.}
With coexisting radar signals, the DMRS symbols $\specRxRawDMRS$ can deviate from $\specRxCommDMRS$, and thus directly estimating CSI by $\specCSIRawEst = \specRxRawDMRS \oslash \specTxCommDMRS$ can be erroneous.
To overcome this, we enhance the DMRS symbol-based CSI estimation to jointly estimate CSI from multiple adjacent DMRS symbols.
Specifically, we employ the Hampel filter~\cite{hampel1974influence} over DMRS symbols, which is repeated for $\symNum$ times for each subcarrier, respectively.
This Hampel filter identifies the interfered DMRS symbols by computing the median absolute deviation (MAD) within a sliding window of multiple estimated CSIs and replaces the outliers with the median of the remaining CSIs.
Here, the median is calculated on the amplitude and phase of the complex-valued CSI.
In {\name}, we empirically take the sliding window length as three DMRS symbols for the Hampel filter, which is the smallest value to provide a meaningful median.
We denote the output of the Hampel filter as the interference-free CSI, $\specCSICommEst$.
Fig.~\ref{fig: radar-separation}(\emph{top}) shows an example of CSI estimation from a correct and an interfered DMRS symbol.
The experiments were conducted using our over-the-air testbed, described in \S\ref{ssec: implementation-testbed}.
It can be seen that certain subcarriers in the CSI estimated from one DMRS symbol deviate from those obtained from the others, due to interference caused by a radar pulse.
This deviation can be effectively mitigated after applying the Hampel filter.

\myparatight{Uplink 5G signal reconstruction and cancellation.}
The uplink CSI $\specCSICommEst$ after the Hampel filter is applied for the data symbols nearby, which is $\specTxRawDataEst = \specRxRawData \oslash \specCSICommEst$.
We assume that the interfered uplink data symbols can still be correctly demodulated (or corrected by the following decoding process).
Since the modulation order is known at the BS, the transmitted constellation points without noise and interference are given and therefore, the transmitted data symbol $\specTxCommDataEst$ can be inferred by rounding to the nearest constellation points:
\begin{align}
    \specTxCommDataEst = \round{\specTxRawDataEst} = \round{\specRxRawData \oslash \specCSICommEst}.
    \label{eq: transmitted-data-symbol-estimation}
\end{align}
The error vector driving $\specTxRawDataEst$ away from constellation points $\specTxCommDataEst$ comes from the noises $\gridRxNoise$ and the radar signal interference $\gridRxRadar$.
An experimental example of this process is illustrated in Fig.~\ref{fig: radar-separation}(\emph{bottom}), where the data symbol interfered by the radar signals (red) has a larger error vector than the uninterfered data symbol (green). This error vector is mainly responsible for the interference from the radar signals.

Finally, the noise- and interference-free uplink signal $\specRxCommEst$ can be reconstructed based on the estimated CSI $\specCSICommEst$ and transmitted signals, which are branched into the data symbols $\specTxCommDataEst$ given by {\eqref{eq: transmitted-data-symbol-estimation}}, and the DMRS symbols directed from the pre-defined $\specTxCommDMRS$.
This reconstruction can be formulated respectively for the data and DMRS symbols as
\begin{align}
    \begin{cases}
        \specRxCommDataEst &= \specCSICommEst \odot \specTxCommDataEst \\
        \specRxCommDMRSEst &= \specCSICommEst \odot \specTxCommDMRS
    \end{cases}
    \quad \Rightarrow \quad \specRxCommEst.
\end{align}
The output of this module, i.e., residual symbols $\specRxRes$ (or resource grids $\gridRxRes$), can be acquired by subtracting the reconstructed received symbols $\specRxCommEst$ (or resource grids $\gridRxCommEst$):
\begin{align}
    \specRxRes = \specRxRaw - \specRxCommEst \quad \Rightarrow \quad
    \gridRxRes = \gridRxRaw - \gridRxCommEst.
\end{align}

If the uplink 5G transmission is interfered by radar signals, $\gridRxRes$ is expected to retain information about the interfering radar signal, whose parameters can then be extracted by the subsequent two modules of {\name}.
Otherwise, $\gridRxRes$ remains close to the noise floor.

\subsection{Resource Grid Reshaping}
\label{ssec: design-reshaping}

As shown in Fig.~\ref{fig:system-overview}, this \emph{resource grid reshaping} module adjusts the frequency and time resolution of $\gridRxRes$ into $\gridRxPoolSet$ to incorporate the actual radar pulse durations and bandwidths, and reduces the computation cost in the subsequent template correlation module.
In the time domain, the radar pulse duration (e.g., {0.5--2.5}\thinspace{\usec} for P0N\#1, and {3--5}\thinspace{\usec} for Q3N\#1) can be much shorter than an OFDM symbol ($\symTime={33.3}\thinspace\textrm{\usec}$), making them difficult to distinguish.
In the frequency domain, the radar pulse bandwidth varies in a large range (e.g., {1}\thinspace{MHz} for P0N\#1 and P0N\#2, and up to {100}\thinspace{MHz} for Q3N\#1 and Q3N\#3), whose expected frequency resolution thereby varies as well, but is overall much larger than that provided by the 5G resource grid ($\freqSub={30}\thinspace\textrm{kHz}$).

\myparatight{Improving the time resolution by re-FFT.}
We increase the time resolution of the $\gridRxRaw$ by an integer factor of $\refftNum \in \mathbb{N}$, i.e., from $\symTime$ to $\symTime/\refftNum$, while decreasing its frequency resolution by $\refftNum$, i.e., from $\freqSub$ to $\refftNum \cdot \freqSub$.
First, an $\subNum$-point IFFT is conducted on each OFDM symbol, $\specRxRes$, that converts the $\subNum$ frequency-domain symbols across subcarriers to the $\subNum$ time-domain I/Q samples.
Then, the $\subNum$ time-domain I/Q samples are split into $\refftNum$ non-overlapped equal-length segments, each containing $\subNum/\refftNum$ I/Q samples.
Finally, an $\left( \frac{\subNum}{\refftNum} \right)$-point FFT is conducted for $\refftNum$ times respectively on the $\refftNum$ segmented I/Q samples back to the frequency domain.
By concatenating the FFT outputs across the $\symNum$ OFDM symbols in $\gridRxRes$, we obtain the re-FFT-ed resource grid, denoted as $\gridRxFFT \in \mathbb{C}^{\left(\frac{\subNum}{\refftNum}\right) \times \left(\refftNum\symNum\right)}$.

\begin{table}[!t]
    \centering
    \includegraphics[width=0.9\columnwidth]{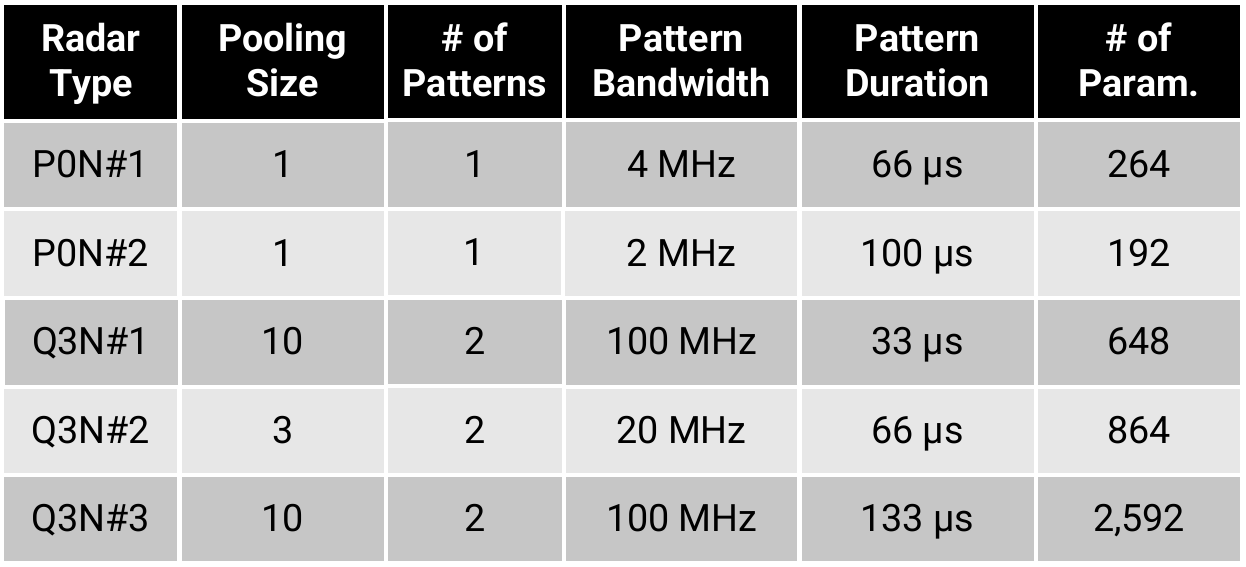}
    \caption{The hyper-parameter setting of the five radar types.}
    \vspace{-8mm}
    \label{fig: hyper-parameter}
\end{table}

\myparatight{Subcarrier maximum pooling.}
The frequency resolution of the re-FFT-ed resource grids $\gridRxFFT$ may still be unnecessarily high (e.g., {120\thinspace{kHz}} per subcarrier with $\refftNum=4$).
Therefore, we conduct max pool every $\poolSize$ adjacent subcarriers into one by their maximum amplitude, which shrinks the frequency dimension of $\gridRxFFT$ by a factor of $\poolSize$.
Notably, the frequency resolution requirements vary significantly across radar types, whose bandwidth range can either be very small (e.g., P0N\#1, P0N\#2, and Q3N\#2) or large (e.g., Q3N\#1 and Q3N\#3).
Therefore, we consider different pooling sizes for each radar type, and define $\poolSize^{(\radarIdx)}$ as the pooling size for the $\radarIdx$-th radar type. Generally, a radar type of small bandwidth is assigned a small $\poolSize^{(\radarIdx)}$ for a finer-grained frequency resolution but a smaller compression ratio, and vice versa.
The selection of the pooling size $\poolSize^{(\radarIdx)}$ in {\name} is detailed in Table~\ref{fig: hyper-parameter}.
In this way, the output of the subcarrier max pooling is formulated as a set of resource grids $\gridRxPoolSet = \left \{ \gridRxPool^{(\radarIdx)} \right \}$, where each resource grid $\gridRxPool^{(\radarIdx)} \in \mathbb{R}^{\subNum^{(\radarIdx)} \times \refftNum\symNum}$ has the identical dimension on time ($\symNum$) while the customized dimensions on the frequency ($\subNum^{(\radarIdx)} = \subNum / \left(\refftNum \cdot \poolSize^{(\radarIdx)}\right)$).
Combining the two reshaping operations, the new frequency resolution $\freqSub^{(\radarIdx)}$ and the new time resolution $\symTime^{\prime}$ of the $\radarIdx$-th resource grid $\gridRxPool^{(\radarIdx)}$ are given by
\begin{align}
    \freqSub^{(\radarIdx)} = \left(\refftNum \cdot \poolSize^{(\radarIdx)}\right) \cdot \freqSub,
    \quad \symTime^{\prime} =  \frac{1}{\refftNum} \cdot \symTime.
\end{align}

\subsection{Radar Template Correlation}
\label{ssec: design-correlation}

\begin{figure}[!t]
    \centering
    \includegraphics[width=0.98\columnwidth]{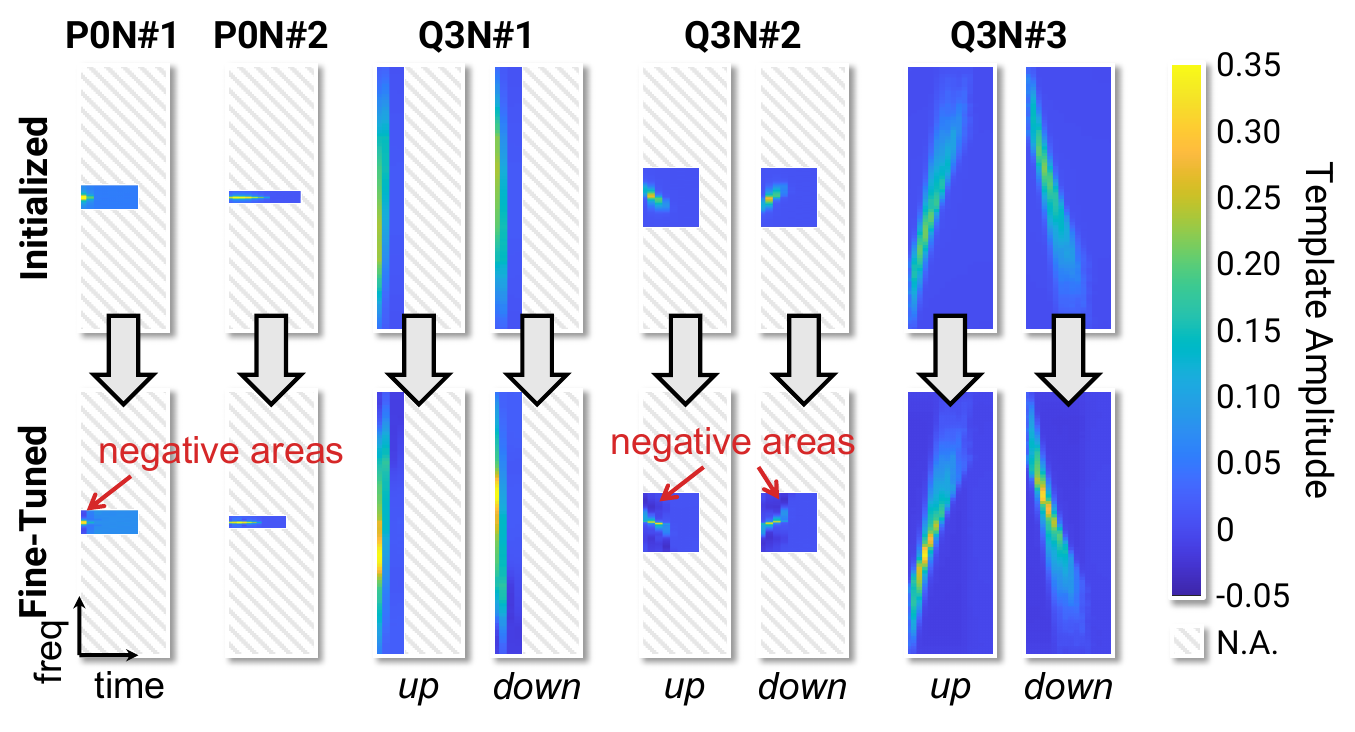}
    \vspace{-3mm}
    \caption{Based on a synthetic dataset, the initialized (by averaging the radar patterns) and fine-tuned (by minimizing the cross-entropy loss) radar templates of the five radar types.}
    \vspace{-5mm}
    \label{fig:template-example}
\end{figure}

Finally, the \emph{radar template correlation} module detects, classifies, and localizes the radar signals by respectively correlating the set of pooled resource grids $\gridRxPoolSet$ with a set of radar templates $\templateSet$, as shown in Fig.~\ref{fig:system-overview}.
Different radar types have different durations in time and bandwidths in frequency, leading to distinct patterns on the resource grids $\gridRxRaw$ or $\gridRxPoolSet$.
Motivated by this observation, we proposed correlating a radar signal ``template'' that captures such patterns with the residual resource grids to effectively detect and localize the radar pulses during each time slot.
This \emph{radar template correlation} module performs correlation of the set between the pooled resource grids ($\gridRxPoolSet$) with a set of radar templates ($\templateSet$) to detect, classify, and localize radar signals.
This process is equivalent to a single-convolution-layer model, which is more computationally efficient compared to more complex ML models such as those based on YOLO~\cite{sarkar2021deepradar, sarkar2024radyololet, ghosh2024sparc}.
More importantly, we demonstrate that these radar templates can be directly constructed from \emph{purely synthetic datasets} without requiring any experimental data, therefore enabling true zero-shot template generation.

\myparatight{Template correlation modeling.}
We consider a set of templates, denoted as $\templateSet = \left \{ \template^{(\radarIdx)} \right \}$, where the template corresponding to the $\radarIdx$-th radar type, $\template^{(\radarIdx)} \in \mathbb{R}^{\layerNum^{(\radarIdx)} \times \subNum^{(\radarIdx)} \times \symNum^{(\radarIdx)}}$, is a real-valued 3D tensor.
The first channel dimension, $\layerNum^{(\radarIdx)}$, refers to the number of patterns the $\radarIdx$-th radar type have, which is $\layerNum^{(\radarIdx)}=1$ for the tone-based radar types (P0N\#1 and P0N\#2), and $\layerNum^{(\radarIdx)}=2$ for the chirp-based radar types (Q3N\#1, Q3N\#2, and Q3N\#3): one for the up-chirps and one for the down-chirps.
The second dimension $\subNum^{(\radarIdx)}$ represents the pattern's frequency-domain bandwidth (i.e., $\subNum^{(\radarIdx)} \cdot \freqSub^{(\radarIdx)}$), and the third dimension $\symNum^{(\radarIdx)}$ represents the time-domain duration (i.e., $\symNum^{(\radarIdx)} \cdot \symTime^{\prime}$).
These two dimensions, $\subNum^{(\radarIdx)}$ and $\symNum^{(\radarIdx)}$, vary over different radar types, and are set to slightly larger than the actual bandwidth/duration of the $\radarIdx$-th radar type, as detailed in Table~\ref{fig: hyper-parameter}.
For the $\radarIdx$-th radar type, the template correlation takes place between the pooled resource grid $\gridRxPool^{(\radarIdx)}$ and the template $\template^{(\radarIdx)}$.
This correlation operation can be formulated as a single 2D convolutional layer with one input channel, $\layerNum^{(\radarIdx)}$ output channels, and a kernel size of $\subNum^{(\radarIdx)} \times \symNum^{(\radarIdx)}$, followed by a max pooling layer over the $\layerNum^{(\radarIdx)}$ dimension.
We further pad zeros to $\gridRxPool^{(\radarIdx)}$ so that the correlation results, denoted as $\gridRxOut^{(\radarIdx)} \in \mathbb{R}^{\subNum^{(\radarIdx)} \times \symNum^{(\radarIdx)}}$, have the same dimension as $\gridRxPool^{(\radarIdx)}$.
To sum up, this correlation process from $\gridRxPool^{(\radarIdx)}$ to $\gridRxOut^{(\radarIdx)}$ can be written as
\begin{align}
    \gridRxOut^{(\radarIdx)}[\subIdx, \symIdx] &= \max_{\layerIdx} \sum_{i = -\frac{\symNum^{(\radarIdx)}}{2}+1}^{\frac{\symNum^{(\radarIdx)}}{2}} \sum_{j=1}^{\subNum^{(\radarIdx)}} \template^{(\radarIdx)}[\layerIdx, i, j] \cdot \gridRxPool^{(\radarIdx)}[\subIdx+i, \symIdx+j].
    \label{eq: template-correlation-process}
\end{align}
Essentially, $\gridRxOut^{(\radarIdx)}[\subIdx, \symIdx]$ indicates the likelihood of the $\radarIdx$-th radar type's existence at a center frequency of $\subIdx \cdot \freqSub^{(\radarIdx)}$ and a starting time of $\symIdx \cdot \symTime^{\prime}$, i.e., during the $\symIdx$-th symbol.
The radar sensing tasks are then equivalent to finding
\begin{align}
\hspace{-3mm}
    \gridRxOpt = \max_{\radarIdx, \subIdx, \symIdx} \gridRxOut^{(\radarIdx)}[\subIdx, \symIdx],\
    \left\{ \radarIdxOpt, \subIdxOpt, \symIdxOpt \right\} = \arg \max_{\radarIdx, \subIdx, \symIdx} \gridRxOut^{(\radarIdx)}[\subIdx, \symIdx].
\end{align}
Finally, {\name} returns the detection ($\outputDetectEst$), classification ($\outputClassEst$), and localization ($\outputLocalFreqEst$ and $\outputLocalTimeEst$) results given by
\begin{align}
    \outputDetectEst &=
    \begin{cases}
        1, &\text{for}~\gridRxOpt \geq \outputDetectThres \\
        0, &\text{for}~\gridRxOpt < \outputDetectThres
    \end{cases} 
    \label{eq: radar-sensing-detection}
    \\
    \outputClassEst &= \radarIdxOpt, ~\outputLocalFreqEst = \subIdxOpt \cdot \freqSub^{(\radarIdxOpt)}, ~\outputLocalTimeEst = \symIdxOpt \cdot \symTime^{\prime},
    \label{eq: radar-sensing-classidfication-localization}
\end{align}
where $\outputDetectThres \in \mathbb{R}$ is a pre-defined threshold used to determine the existence of radar signals.
Note that $\outputDetectEst=0$ means there is no radar pulse detected on the given resource grid $\gridRxRaw$, and thereby the other outputs, $\outputClassEst$, $\outputLocalFreqEst$, and $\outputLocalTimeEst$, become invalid.

\myparatight{Multiple radar pulse appearance.}
Our template correlation model can be easily extended to scenarios where multiple radar pulses appear on the same resource grid, $\gridRxRaw$.
Specifically, the multiple local maximums $\gridRxOut^{(\radarIdx)}[\subIdx, \symIdx]$ satisfying $\gridRxOut^{(\radarIdx)}[\subIdx, \symIdx] > \outputDetectThres$ in {\eqref{eq: radar-sensing-detection}} indicate the detection of multiple radar pulses within $\gridRxRaw$. Their respective $\outputClassEst$, $\outputLocalFreqEst$ and $\outputLocalTimeEst$ can be given by plugging $\gridRxOut^{(\radarIdx)}[\subIdx, \symIdx]$ in {\eqref{eq: radar-sensing-classidfication-localization}}.

\myparatight{Benefits of the template correlation model.}
This template correlation model in \eqref{eq: template-correlation-process} only contains a single 2D convolutional layer with one or two channels, making it far more lightweight than existing ML models (e.g., based on YOLO) in terms of the number of parameters, therefore incurring minimum computation cost on the BS.
Moreover, the template correlation model is strictly linear, with no nonlinear activation functions.
This means that its outputs remain invariant under additional gain or loss applied to the receive resource grids, ensuring portability across receiver hardware (with the exception of $\outputDetectEst$, which depends on the specific $\outputDetectThres$, whose selection is described in \S\ref{ssec: evaluation-detection}).

\myparatight{Template initialization by pattern averaging.}
Ideally, the template $\template^{(\radarIdx)}$ should imitate the pattern of the $\radarIdx$-th radar type appeared on the pooled resource grid $\gridRxPool^{(\radarIdx)}$.
Given a radar type, the radar pulse's bandwidth and duration can vary in a small range, causing the radar pattern to vary slightly accordingly.
In {\name}'s zero-shot template generation, we develop a synthetic dataset of the five radar signal types (see details in \S\ref{ssec: implementation-dataset}), including the time-domain waveforms of multiple radar pulse samples with randomized bandwidths and durations.
Their appearing patterns on $\gridRxPool^{(\radarIdx)}$ can be simulated following the process ahead.
We initialize each channel in the template $\template^{(\radarIdx)}$ by element-wise averaging the patterns over all the radar pulse samples in the synthetic dataset.
Note that for the three chirp-based radar types, the two channels are initialized by the up-chirps and down-chirps separately.
Finally, we normalize the power of these initialized templates to the same.
Fig.~\ref{fig:template-example} shows the zero-shot initialized templates by pattern averaging, requiring no experimental measurements.

\myparatight{Template fine-tuning by cross-entropy loss.}
The zero-shot initialized templates achieve good performance on the existence detection ($\outputDetectEst$) and center frequency/starting time localization ($\outputLocalFreqEst$ and $\outputLocalTimeEst$) tasks, while being unsatisfactory on the radar type classification ($\outputClassEst$) task.
To further improve the classification accuracy, we conduct zero-shot fine-tuning for the templates $\templateSet$ on the same synthetic dataset. This is achieved by minimizing the cross-entropy loss between the predicted radar type $\outputClassEst$ and the ground truth radar type $\outputClass$ for each radar pulse sample.
A comparison of the initialized templates and the zero-shot fine-tuned templates is shown in Fig.~\ref{fig:template-example}.
Interestingly, the zero-shot fine-tuned templates of the two narrowband radar types, P3N\#1 and Q3N\#2, have \emph{negative areas} on the top/bottom in the frequency dimension.
This means radar signals with a pattern covering both negative areas have a wide bandwidth, and are less likely to be classified as these two narrowband radar types.
\section{Implementation}
\label{sec:implementation}

\subsection{Over-the-Air (OTA) 5G Link}
\label{ssec: implementation-testbed}

\begin{figure}[!t]
    \centering
    \includegraphics[width=0.99\columnwidth]{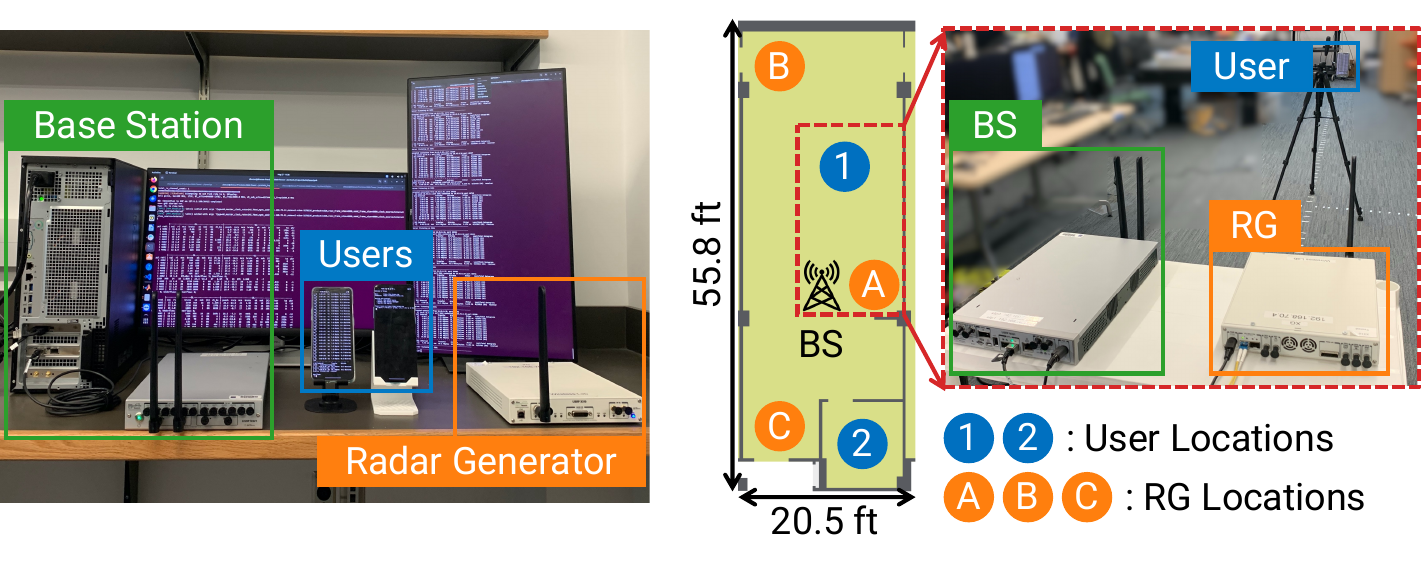}
    \vspace{-3mm}
    \caption{OTA testbed including the BS, users, and RG (\emph{left}), and setup for collecting the two experimental datasets (\emph{right}).}
    \vspace{-3mm}
    \label{fig: experiment-setup}
\end{figure}

\myparatight{5G core network and radio access network.}
We employ Open5GS (\texttt{commit-6d80d4322})~\cite{open5gs}, an open-source 5G NR core network software platform compliant with 3GPP Release-17~\cite{3gpprelease17}.
Specifically, the Open5GS is deployed and configured as a standalone 5G core to connect with the 5G BS.
Additionally, we register OnePlus 8T as the 5G users in the subscriber list using the Open5GS WebUI and MongoDB database.
We use the open-source 5G radio access network (RAN) stack, srsRAN version 24.04.0 (\texttt{commit-51e44a642}).

\myparatight{5G configuration.}
The established OTA 5G link between the BS and user(s) is centered at {3.60}\thinspace{GHz} (CBRS band) with a channel bandwidth of {100}\thinspace{MHz}.
This band follows TDD with a transmission periodicity of {10} time slots lasting for {5}\thinspace{ms}, where up to {5} uplink slots can be allocated.
We follow 5G NU1 with $\freqSub={30}\thinspace\textrm{kHz}$ and an FFT size of $\fftSize={4,096}$, which corresponds to a subcarrier number of $\subNum={3,276}$ within the {100}\thinspace{MHz} channel bandwidth.

\begin{figure}[!t]
    \begin{minipage}{0.47\columnwidth}
        \centering
        \includegraphics[width=1\linewidth]{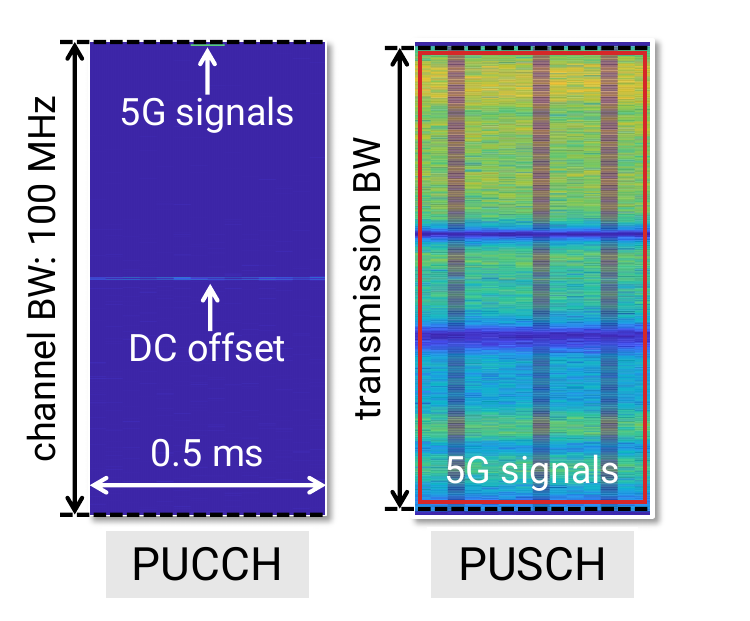}
        \vspace{-7mm}
        \caption{The resource grid examples of the 5G signals in PUCCH and PUSCH.}
        \vspace{-3mm}
        \label{fig: demo-5g-signal}
    \end{minipage}
    \hspace{0.08in}
    \begin{minipage}{0.47\columnwidth}
        \centering
        \includegraphics[width=1\linewidth]{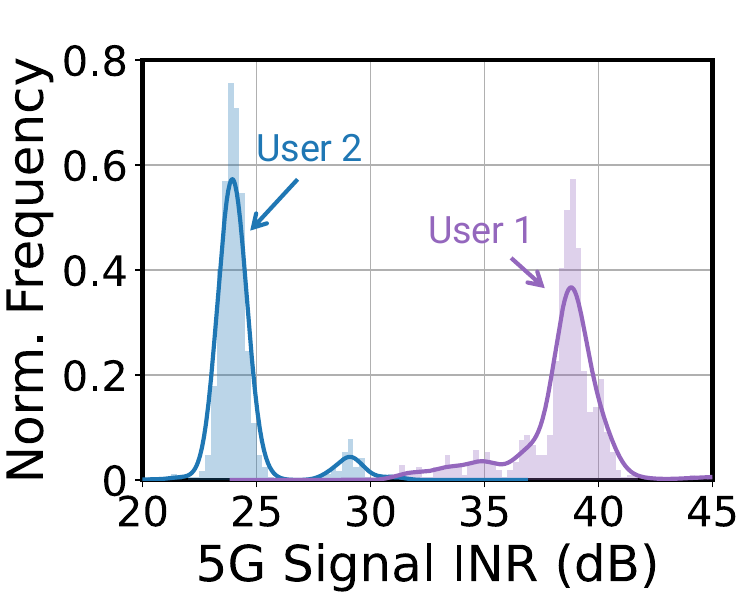}
        \vspace{-6mm}
        \caption{The 5G signal INR distribution from two users in the PUSCH dataset.}
        \vspace{-3mm}
        \label{fig: snr-hist}
    \end{minipage}
\end{figure}

\myparatight{USRP N321 as the BS.}
We use USRP N321~\cite{usrpN321}, a high-performance SDR, as the 5G BS.
We configure its sampling rate to {122.88}\thinspace{MHz} to comply with the 5G configuration.

\myparatight{Commodity smartphones as the users.}
OnePlus 8T phones with Android~11 are used to connect to our 5G standalone platform.
The phone is equipped with the SIM card, SysmoISIM-SJA5, which is reprogrammed by a Python software suite, pySim, in order to support the CBRS band.
Moreover, Force 5G Only~\cite{5gSwitchApp} is installed on each user to force the user to connect exclusively to 5G networks.
During the experiments, the 5G UDP traffic between the BS and users is created by running iPerf3~\cite{Iperf3} on the users.
As shown in Fig.~\ref{fig: demo-5g-signal}, the 5G traffic can be categorized into PUCCH (with little bandwidth occupancy for the control information) or PUSCHs (with large bandwidth occupancy for uplink data), and the subcarrier allocations are automatically determined according to the users' communication requirements.
Since the users' transmitted power cannot be controlled, we tune the 5G signal SNR (or INR for {\name}'s radar sensing scenario) by different BS-user distances. 
We consider two user locations, as shown in Fig.~\ref{fig: experiment-setup}. resulting in INR of (approximately) {24.33--38.38}\thinspace{dB} during PUSCH, as shown in Fig.~\ref{fig: snr-hist}.
Note that this INR variance comes from not only the different user locations, but also the different number of subcarriers allocated.

\subsection{Radar Dataset Generation}
\label{ssec: implementation-dataset}

We construct one synthetic dataset (split into training and testing) and two OTA experimental datasets (testing only) collected under 5G PUCCH and PUSCH, respectively.
All datasets share a consistent format: each sample consists of a complex-valued matrix $\gridRxRaw \in \mathbb{C}^{3,276 \times 14}$ spanning {14} OFDM symbols, each with {3,276} subcarriers, along with ground truth labels for radar detection ($\outputDetect$), classification ($\outputClass$), as well as frequency and time localization ($\outputLocalFreq$ and $\outputLocalTime$).
Due to the large inter-pulse duration of radar signals, each resource grid contains at most one radar pulse (see also Table~\ref{fig:radar-characterstics}).

\myparatight{Radar waveform generation.}
We leverage the NIST’s simulated radar waveform generator~\cite{caromi2019rf, nistgithub} (\texttt{commit-b7ba94b}), a software tool built in MATLAB to generate time-domain radar waveforms of the five radar types.
For each radar type, we generate {1,000} radar waveforms at a sampling rate of {100}\thinspace{MHz} with varying bandwidths, durations, and chirp directions (if applicable).
Each waveform is a time-domain I/Q sample sequence represented by a complex-valued vector.
The generated waveforms are randomly split into {9:1} for training and testing datasets.

\myparatight{Synthetic datasets for template training and testing.}
The synthetic dataset is generated by numerically simulating the resource grid $\gridRxRaw$ with 5G signals interfered by radar signals.
Note that we also include the case when there are no radar signals, which is labeled with $\outputDetect = 0$.
In the simulation, we consider a frequency-flat channel with additive white Gaussian noise at varying SNRs.
We randomize the center frequency of the radar signals uniformly located from {$-$40}\thinspace{MHz} to {$+$40}\thinspace{MHz}, and the starting time uniformly within the time slot with a duration of {0.5}\thinspace{ms}.
The entire synthetic dataset is randomly split into {9:1} for training and testing purposes.
Note that this training dataset is for zero-shot radar template initialization and fine-tuning as described in \S\ref{ssec: design-correlation}.
Specifically for fine-tuning, we employ stochastic gradient descent (SGD)~\cite{bottou2010large} with a learning rate of $10^{-3}$ and a weight decay of $10^{-3}$ as the optimizer for {200} epochs, and pick the best set of templates according to the performance on the synthetic testing dataset.

\myparatight{Experimental datasets (PUSCH/PUCCH) for testing.}
As shown in Fig.~\ref{fig: experiment-setup}, we conduct OTA experiments to collect the testing dataset involving real-world 5G traffic from the commodity phones.
Specifically, we deploy an additional USRP X310 operating at a sampling rate of {100}\thinspace{MHz} as the radar generator (RG) to transmit the radar waveforms, whose transmitted power is tuned so that the received SNR of the radar signals on the BS varies between {10--35}\thinspace{dB}, which is (approximately) equivalent to an SINR between $-$28.4\thinspace{dB} and {10.7}\thinspace{dB} SINR given the 5G signals' INR ranging at {24.33--38.4}\thinspace{dB}.
A center frequency shift between this RG and the BS is added as $\{ \pm 10, \pm 30 \}\thinspace\textrm{MHz}$ for the three small-bandwidth radar types (P0N\#1, P0N\#2 and Q3N\#2); while for the two large-bandwidth radar types (Q3N\#1 and Q3N\#3), it is always fixed as {0}\thinspace{MHz} to ensure the occupied bandwidth of the radar signals within the CBRS band. The starting time is randomized as $\outputLocalTime \sim \mathcal{U}[0, 0.5]\thinspace\textrm{ms}$, and is labeled manually.
Similarly, we also include samples without radar signal transmissions.
All the collected data samples are split into one PUCCH dataset and one PUSCH dataset based on the function of the time slot.
In the PUCCH dataset, the 5G signals only occupy a small transmission bandwidth for control information, which can be treated as interference-free by the 5G signals; as for the PUSCH dataset, the 5G signals span over a large transmission bandwidth (up to the whole channel bandwidth of {100}\thinspace{MHz}), depending on the uplink data demand on the users.
With true zero-shot generation, the radar templates used by {\name} in the OTA setting require no fine-tuning on these experimental datasets, and can be directly employed for testing.

\begin{figure*}[!t]
    \centering
    \includegraphics[width=2.1\columnwidth]{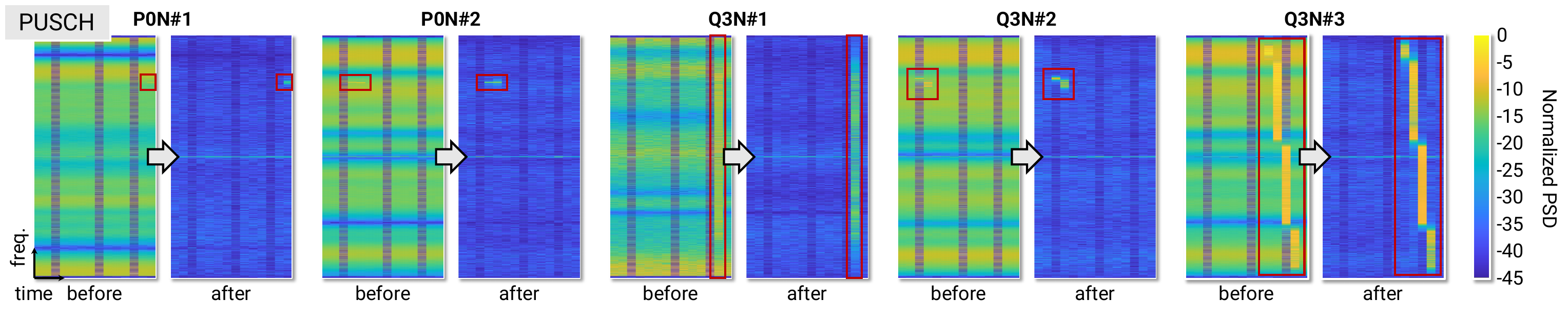}
    \vspace{-10mm}
    \caption{Example PUSCH resource grids with interfering 5G and radar signals before and after radar signal separation.}
    \vspace{-3mm}
    \label{fig:spectrogram-example}
\end{figure*}

\subsection{Baseline Radar Sensing Methods}

We compare {\name}'s performance with three baselines:
energy detection~\cite{urkowitz1967energy}, DeepRadar~\cite{sarkar2021deepradar}, and RadYOLO~\cite{sarkar2024radyololet}.
For fair comparison, DeepRadar and RadYOLO are trained using only the synthetic dataset described in \S\ref{ssec: implementation-dataset}, whose performance on the experimental datasets may degrade.

\myparatight{Energy detection.}
Given a raw resource grid, this method compares the maximum PSD element to a pre-defined threshold to determine the radar existence; the element's position is then used to estimate the radar's center frequency and starting time.
Note that this method only detects radar presence and location; it \emph{does not} classify the radar type.

\myparatight{DeepRadar.}
This is an ML-based method built on YOLO~\cite{redmon2016you} and supports radar detection, classification, and localization. 
The ML model input is the spectrogram in dB scale with customized frequency/time resolution.
We adapt the original DeepRadar model architecture to suit the spectrogram with given dimension, which employs five 2D convolutional layers, followed by three fully-connected layers.

\myparatight{RadYOLO.}
Similar to DeepRadar, RadYOLO adopts the YOLO framework but compresses the frequency dimension of the input spectrogram by {8}$\times$ using a max pooling layer before converting it to the spectrogram in dB scale, which reduces both model size and computational cost. 
Besides, its model architecture employs two ResNet blocks~\cite{he2016deep}.

\section{Evaluation}
\label{sec:evaluation}

\subsection{Radar Signal Separation Examples}

In Fig.~\ref{fig:spectrogram-example}, we illustrate the examples of the experimental resource grids before and after {\name}'s radar signal separation.
These resource grid examples are captured during the PUSCH with heavy uplink traffic.
In particular, the 5G signals occupy a large transmission bandwidth within the entire {100}\thinspace{MHz} channel, which overlaps with the radar signals in both time and frequency.
The 5G signals have an overall INR between {24.3--38.4}\thinspace{dB}, and the five types of radar signals have fixed SNRs at {30}\thinspace{dB}, corresponding to an approximate SINR between $-$8.38\thinspace{dB} and {5.7}\thinspace{dB}.
Across the five radar types, the radar signal separation decreases the INR of the 5G signals by approximately {21.7-32.9}\thinspace{dB}, equivalent to increasing the SINR of the radar signals by the same amount.

\subsection{Radar Existence Detection}
\label{ssec: evaluation-detection}

\begin{figure}[!t]
    \centering
    \includegraphics[width=0.95\columnwidth]{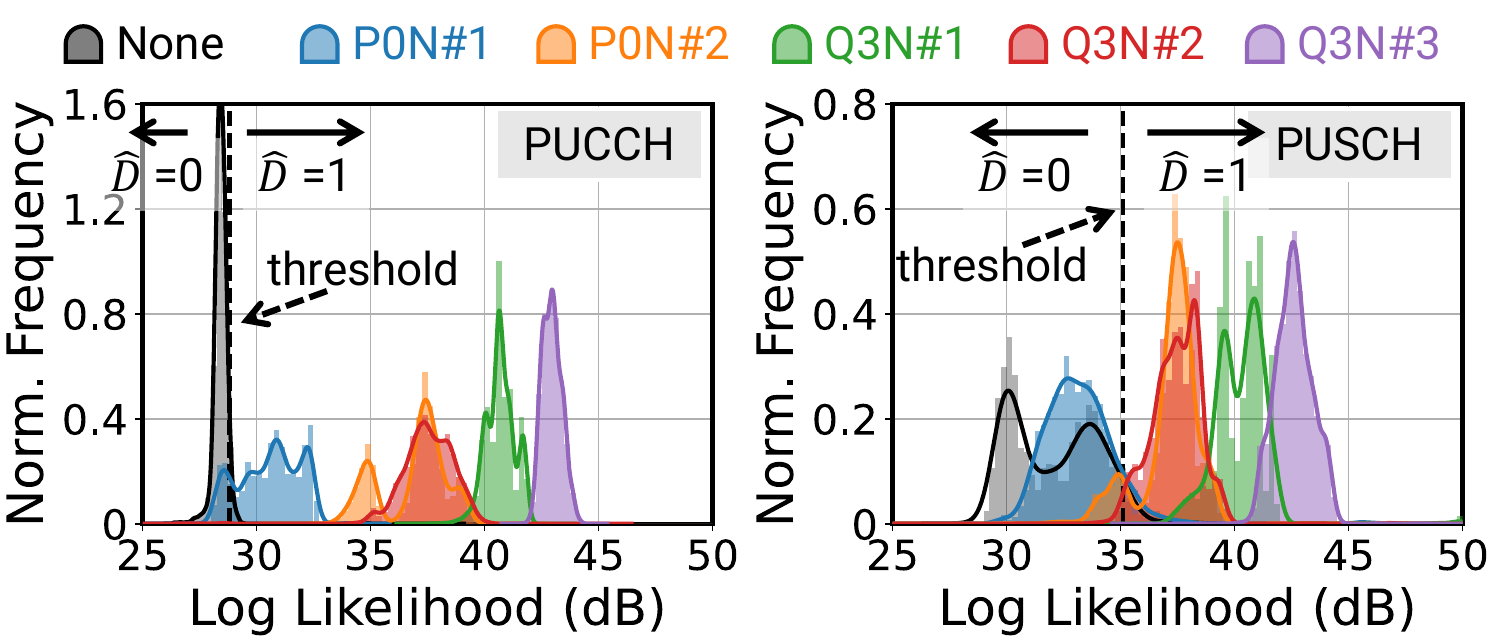}
    \vspace{-3mm}
    \caption{The output likelihood distributions with and without radar signals, and the thresholds for existence detection.}
    \vspace{-3mm}
    \label{fig:detection-hist}
\end{figure}

\myparatight{Likelihood threshold selection.}
In our evaluation, we consider such $\outputDetectThres$ that yields a false alarm probability of {5\%}, and report the corresponding detection probability.
We first examine the distribution of $\gridRxOpt$ with radar signals of {30}\thinspace{dB} SNR or no radar signals, as shown in Fig.~\ref{fig:detection-hist}.
Given the same radar SNR, the radar signals' PSD is fixed, so the total energy of a radar pulse is proportional to its bandwidth and duration, e.g., P0N\#1 has the lowest energy while Q3N\#3 has the highest.
In the fine-tuned radar templates, most values are positive. So, the resource grid with the radar type of a high sum energy tends to yield a high $\gridRxOpt$, while the resource grid without radar signals has the lowest likelihood.
To keep the false alarm probability by {5\%}, $\outputDetectThres$ is set to the {95}-th percentile of $\gridRxOpt$ by the resource grids without radar signals.
Specifically on PUCCH as shown in Fig.~\ref{fig:detection-hist}(\emph{right}), the log threshold $\outputDetectThres$ is set to {28.8}\thinspace{dB}.
Under this $\outputDetectThres$, {97.02\%} of the radar signals can be detected, where the missed cases mainly come from radar type P0N\#1 with the lowest sum energy, whose detection probability is only {85.27\%}.
As for the PUSCH with heavy 5G traffic, the input resource grid contains residual 5G signal energy, which inevitably increases the output likelihood, as shown in Fig.~\ref{fig:detection-hist}(\emph{right}). 
As a result, the threshold $\outputDetectThres$ is increased to {35.1}\thinspace{dB} to maintain the same {5\%} false alarm probability, i.e., {6.3}\thinspace{dB} higher than that on PUCCH.
Hereby, the average detection probability over the five radar types drops to {79.23\%}, and that of radar type P0N\#1 drops to {8.95\%}.

\begin{figure}[!t]
    \centering
    \includegraphics[width=0.95\columnwidth]{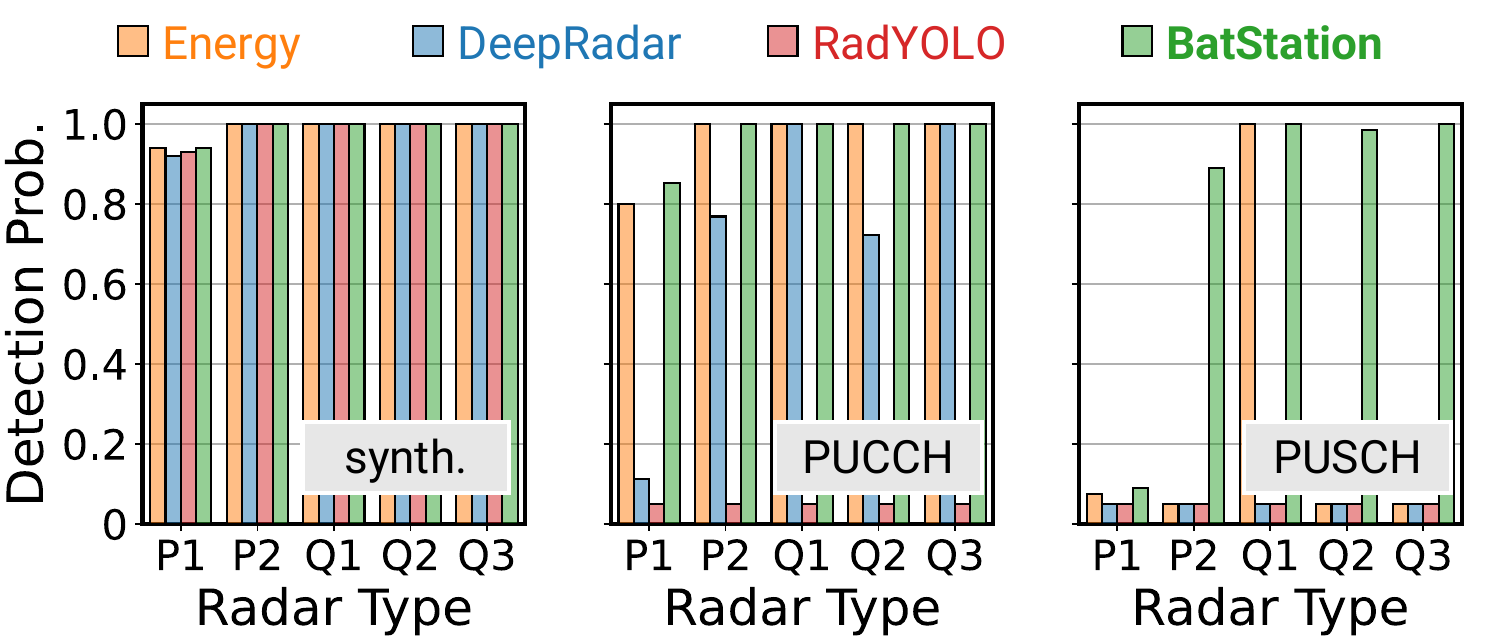}
    \vspace{-3mm}
    \caption{The detection probability comparison of {\namebf} and the three baselines on the three testing datasets.}
    \vspace{-3mm}
    \label{fig:group-bar-detect-new}
\end{figure}

\myparatight{Comparison to baselines.}
Fig.~\ref{fig:group-bar-detect-new} shows the detection probability comparison between {\name} and the three baseline methods over the five radar types on the three testing datasets, where the SNRs of the radar signals are {30}\thinspace{dB}.
On the synthetic dataset, all four methods achieve good detection probability at {98.80/98.38/98.60/98.79\%}, respectively, as shown in Fig.~\ref{fig:group-bar-detect-new}(\emph{left}).
On the experimental PUCCH dataset with moderate interference from the 5G signals, the detection probabilities of {\name} are maintained at {85.27/100/100/99.83/100\%} for P0N\#1, P0N\#2, Q3N\#1, Q3N\#2 and Q3N\#3, as shown in Fig.~\ref{fig:group-bar-detect-new}(\emph{middle}).
On the other hand, the performance of the two ML-based baselines, DeepRadar and RadYOLO, drops significantly for their models converging to the synthetic training datasets, while the baseline, energy detection, remains robust.
As for the PUSCH in Fig.~\ref{fig:group-bar-detect-new}(\emph{right}), {\name} can still detect the last four radar types at detection probabilities of {88.84/100/98.34/100\%}, while that for P0N\#1 drops to {8.95\%}. This is because of the narrow bandwidth and the short time duration of a single P0N\#1 pulse. 
In contrast, all three baselines fail to detect the radar signals during the PUSCH.

\begin{figure}[!t]
    \centering
    \includegraphics[width=0.95\columnwidth]{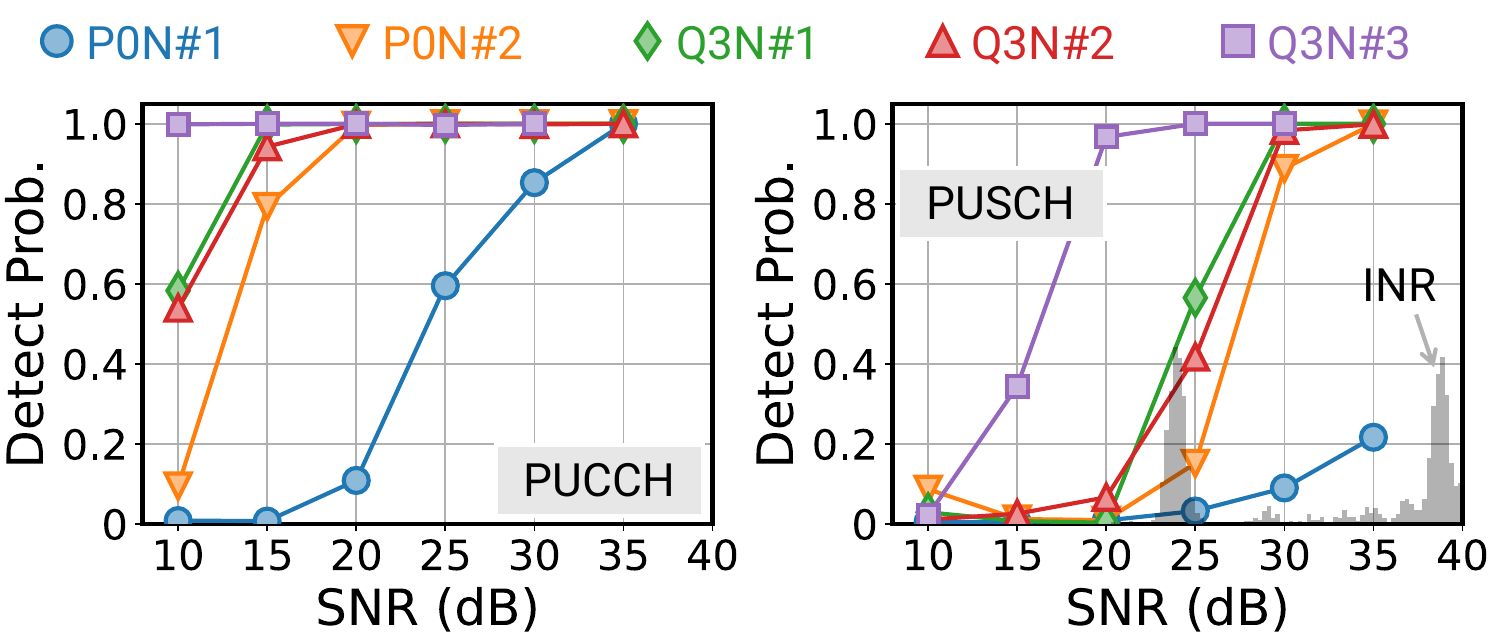}
    \vspace{-3mm}
    \caption{{\namebf}'s detection probability over radar signal SNRs on the PUCCH and PUSCH datasets.}
    \vspace{-3mm}
    \label{fig:line-snr-detect-new}
\end{figure}

\myparatight{Impact of the radar signal SNRs.}
We further examine {\name}'s radar detection over different SNRs.
As shown in Fig.~\ref{fig:line-snr-detect-new}(\emph{left}), the detection probability on PUCCH of radar type Q3N\#3 remains {99.84\%} under {10}\thinspace{dB} SNR. This results from its largest bandwidth and duration, and thereby the highest total sum energy.
The detection probabilities of Q3N\#1 and Q3N\#2 also remain {58.27/53.91\%} under {10}\thinspace{dB} SNR.
As for the PUSCH shown in Fig.~\ref{fig:line-snr-detect-new}(\emph{right}), the detection probability of radar type, Q3N\#3, remains at {96.69\%} under {20}\thinspace{dB} SNR, while it drops to {34.38\%} under {15}\thinspace{dB} SNR.
As for radar types P0N\#2, Q3N\#1, and Q3N\#2, their detection probabilities are {88.84/100/98.34\%} under {30}\thinspace{dB} SNR, and are dropped to {15.09/56.49/41.69\%} under {25}\thinspace{dB}.
Overall, the detection probability tendencies of the five radar types on the PUCCH dataset are close to that on the PUSCH dataset with {15}\thinspace{dB} SNR lower, which indicates the residual 5G signals after the radar signal separation still approximately reduce the radar signals' SINR by {15}\thinspace{dB}, which is already much smaller than its original INR of {24.3--38.4}\thinspace{dB}.

\subsection{Radar Type Classification}

\begin{figure}[!t]
    \centering
    \includegraphics[width=0.95\columnwidth]{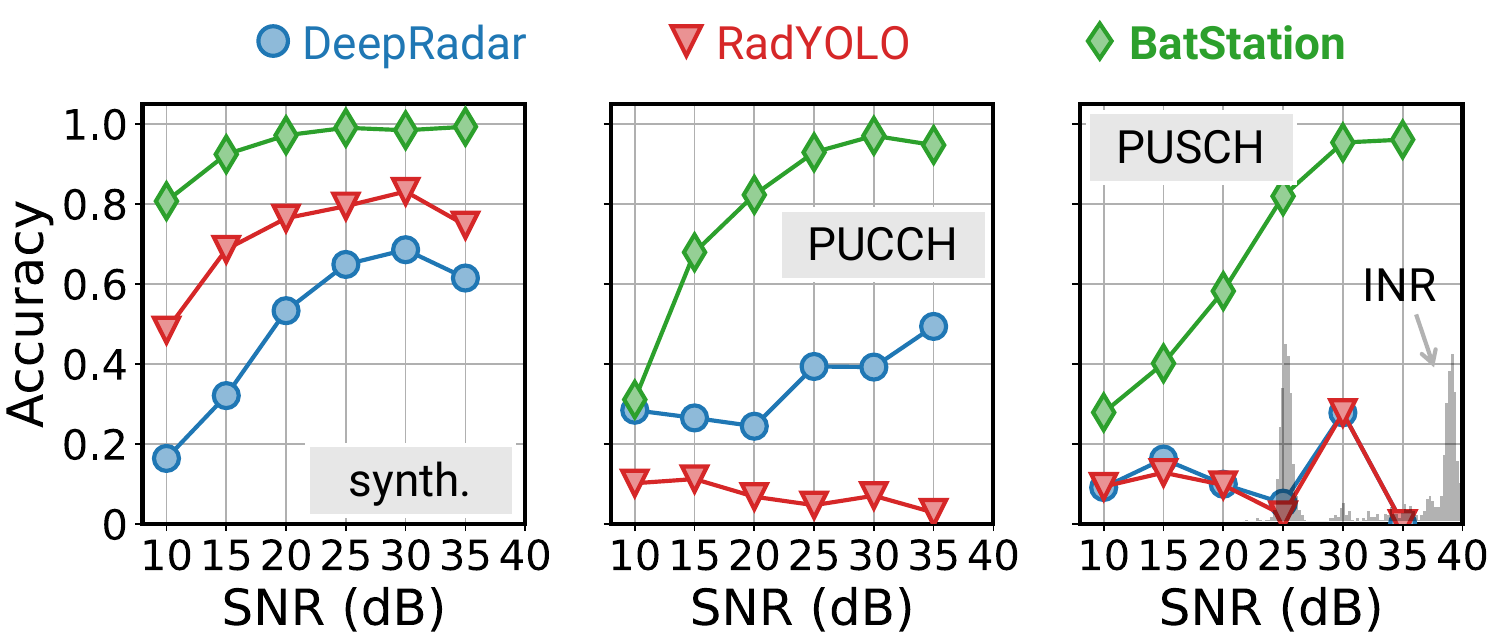}
    \vspace{-3mm}
    \caption{The radar type classification accuracy comparison of {\namebf} and the two ML-driven baselines over SNRs.}
    \vspace{-3mm}
    \label{fig:line-snr-acc}
\end{figure}

\begin{figure}[!t]
    \centering
    \includegraphics[width=0.90\columnwidth]{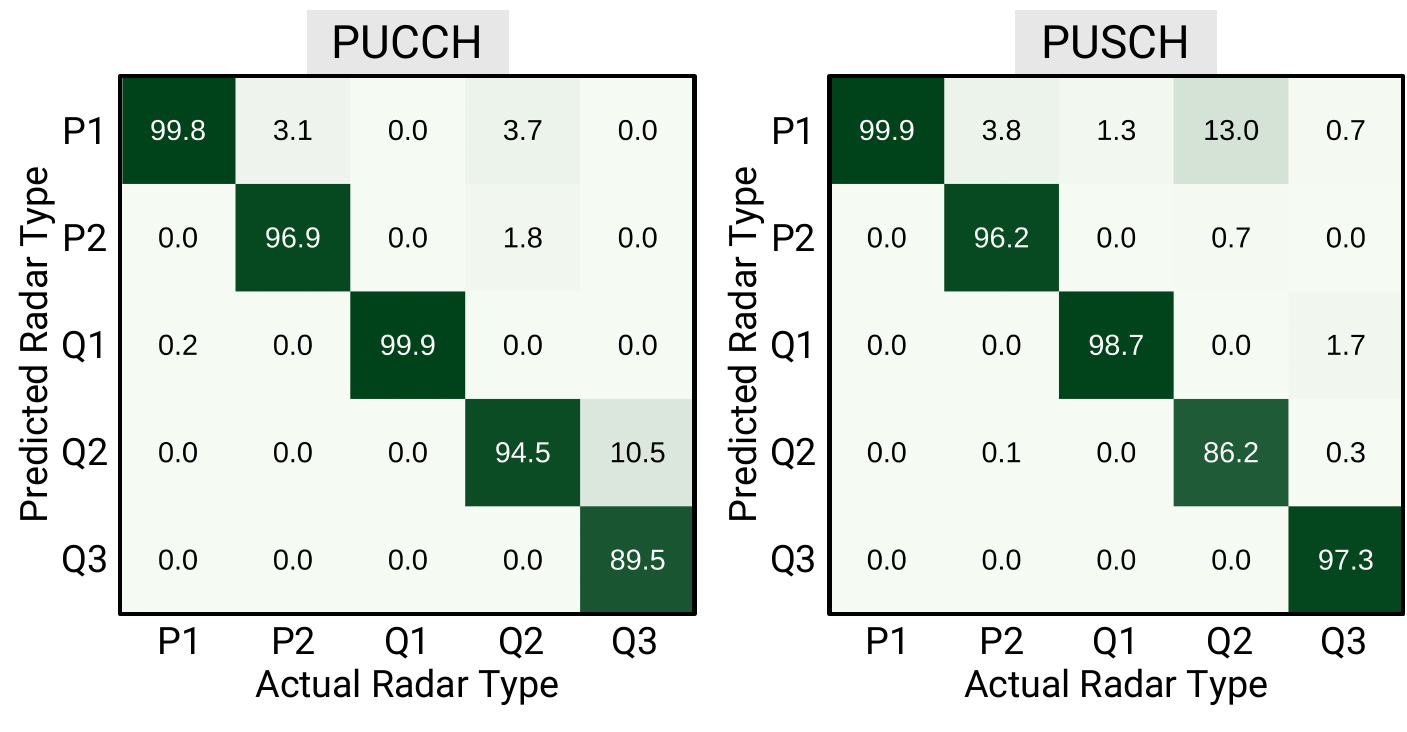}
    \vspace{-3mm}
    \caption{The confusion matrices of {\namebf}'s classification under {30}\thinspace{dB} SNR on the PUCCH and PUSCH datasets.}
    \vspace{-3mm}
    \label{fig:confusion-matrix-new}
\end{figure}

We compare {\name}'s radar type classification performance to the two ML-driven baselines, DeepRadar~\cite{sarkar2021deepradar} and RadYOLO~\cite{sarkar2024radyololet} that have the classification capability.
On the synthetic dataset as shown in Fig.~\ref{fig:line-snr-acc}(\emph{left}), the classification accuracies based on a single radar pulse with {30}\thinspace{dB} SNR reach up to {68.51/83.06\%} for DeepRadar and RadYOLO, respectively, lower than {\name} of {98.37\%}.
However, the classification accuracies cannot exceed {49.33/11.18\%} for the two ML-driven methods on the PUCCH dataset, as shown in Fig.~\ref{fig:line-snr-acc}(\emph{middle}), and {27.73/27.74\%} on the PUSCH dataset, as shown in Fig.~\ref{fig:line-snr-acc}(\emph{right}). Such performance degradation mainly comes from the unseen hardware parameters on the experimental BS hardware (e.g., noise floor, receiving gain).
On the other hand, {\name} can still achieve classification accuracies of {97.00\%} and {95.30\%} under PUCCH and PUSCH datasets under {30}\thinspace{dB} SNR, respectively.
The detailed confusion matrices of {\name}'s five-radar-type classification under {30}\thinspace{dB} SNR are shown in Fig.~\ref{fig:confusion-matrix-new}.
This comparison reveals the hardware portability of {\name}'s template correlation model of {\name}, which can be \emph{directly applied by entire training on the synthetic dataset without any real-world experimental data input}.

\subsection{Frequency/Starting Time Localization}

We normalize the localization error of the center frequency by the channel bandwidth ({100}\thinspace{MHz}), and that of the starting time by the uplink slot duration ({0.5}\thinspace{ms}).
We omit the three baselines for their unsatisfactory classification performance.

\begin{figure}[!t]
    \centering
    \includegraphics[width=0.95\columnwidth]{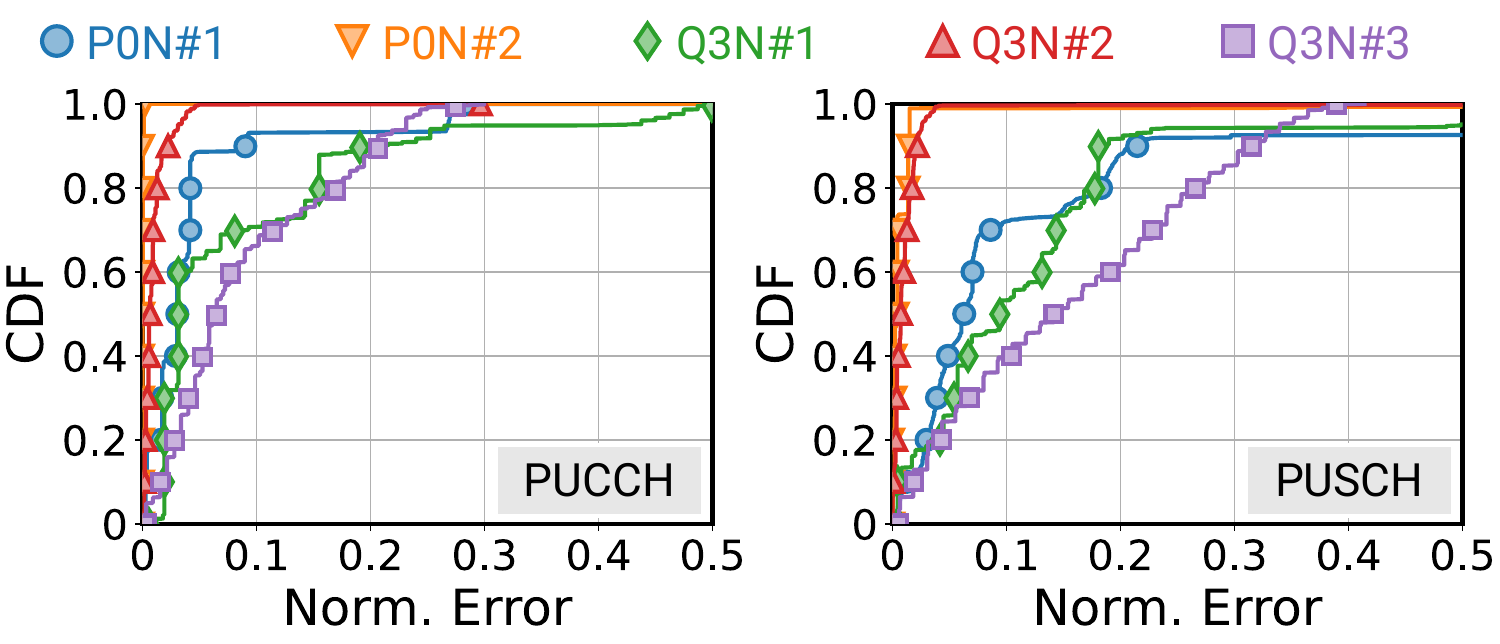}
    \vspace{-3mm}
    \caption{The normalized frequency errors by {\namebf}'s center frequency localization under {30}\thinspace{dB} SNR.}
    \vspace{-3mm}
    \label{fig:error-cdf-freq}
\end{figure}

\myparatight{Center frequency localization.}
Under {30}\thinspace{dB} SNR, the CDFs of the normalized center frequency localization error on the PUCCH and PUSCH datasets is shown in Fig.~\ref{fig:error-cdf-freq}.
Specifically, the median errors for the five radar types are {3.05/0.02/3.15/0.68/6.50\%} on the PUCCH dataset, corresponding to the absolute median error of {3.05/0.02/3.15/0.68/6.50}\thinspace{MHz}; their median errors are {6.29/0.43/9.41/0.73/14.16\%} on the PUSCH dataset, which are {6.29/0.43/9.41/0.73/14.16}\thinspace{MHz}.
Among all five radar types, P0N\#2 and Q3N\#2 have the smallest center frequency error because of their smallest bandwidths. The large errors on Q3N\#1 and Q3N\#3 are because of their largest bandwidths up to {100}\thinspace{MHz}.

\begin{figure}[!t]
    \centering
    \includegraphics[width=0.95\columnwidth]{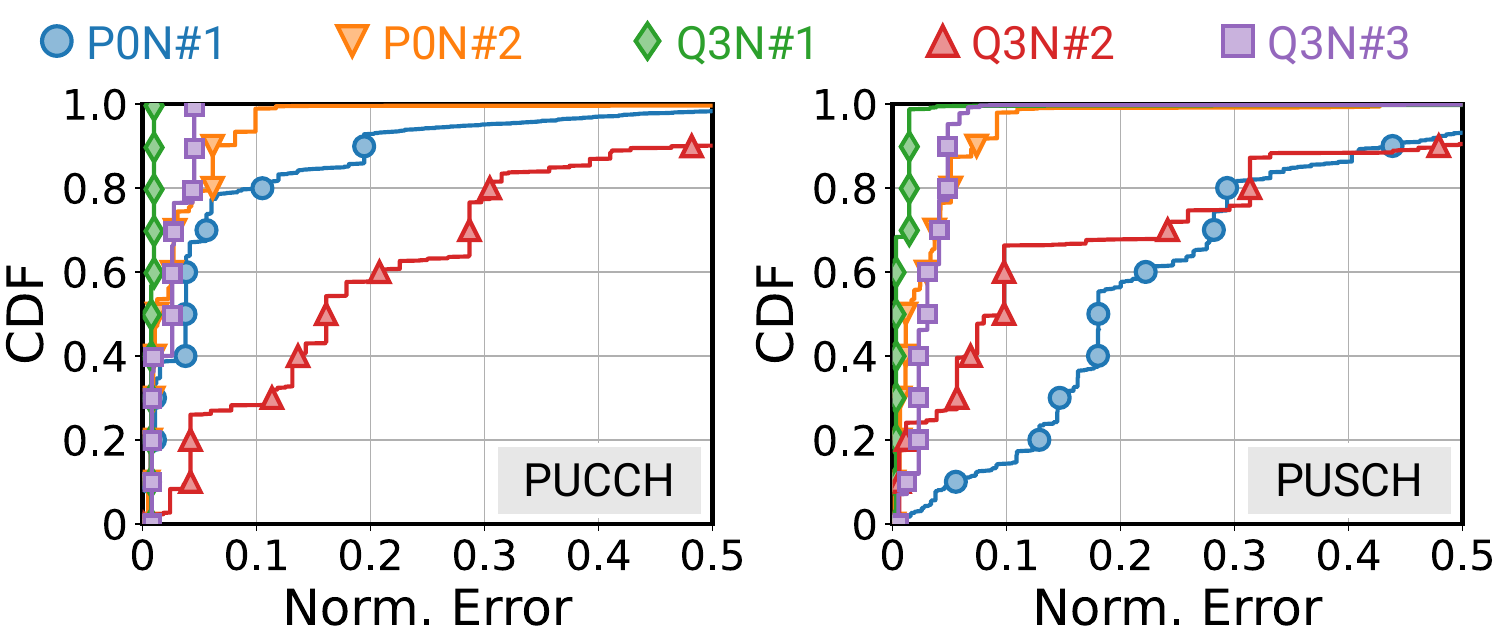}
    \vspace{-3mm}
    \caption{The normalized time errors by {\namebf}'s starting time localization under {30}\thinspace{dB} SNR.}
    \vspace{-5mm}
    \label{fig:error-cdf-time}
\end{figure}

\myparatight{Starting time localization.}
We also examine {\name}'s normalized errors on the starting time localization.
Fig.~\ref{fig:error-cdf-time}(\emph{left}) plots the CDFs of the normalized time errors on the PUCCH dataset, where the median normalized time errors of the five types are {3.77/1.29/0.77/16.11/2.60\%}, corresponding to the absolute errors of {18.85/6.45/3.85/80.56/13.02}\thinspace{\usec}.
In addition, as shown in Fig.~\ref{fig:error-cdf-time}(\emph{right}), the median normalized time errors on the PUSCH dataset are {18.06/1.17/0.31/9.79/3.06\%} for the five radar types, i.e., {90.29/5.84/1.56/48.96/15.28}\thinspace{\usec}.
Specifically, P0N\#2, Q3N\#1 have small starting time errors due to their shortest durations, while the long duration of Q3N\#3 is compensated by its large bandwidth with high energy. On the PUSCH dataset, the performance degradation of P0N\#1 comes from its low detection probability.

\begin{figure}[!t]
    \centering
    \includegraphics[width=0.95\columnwidth]{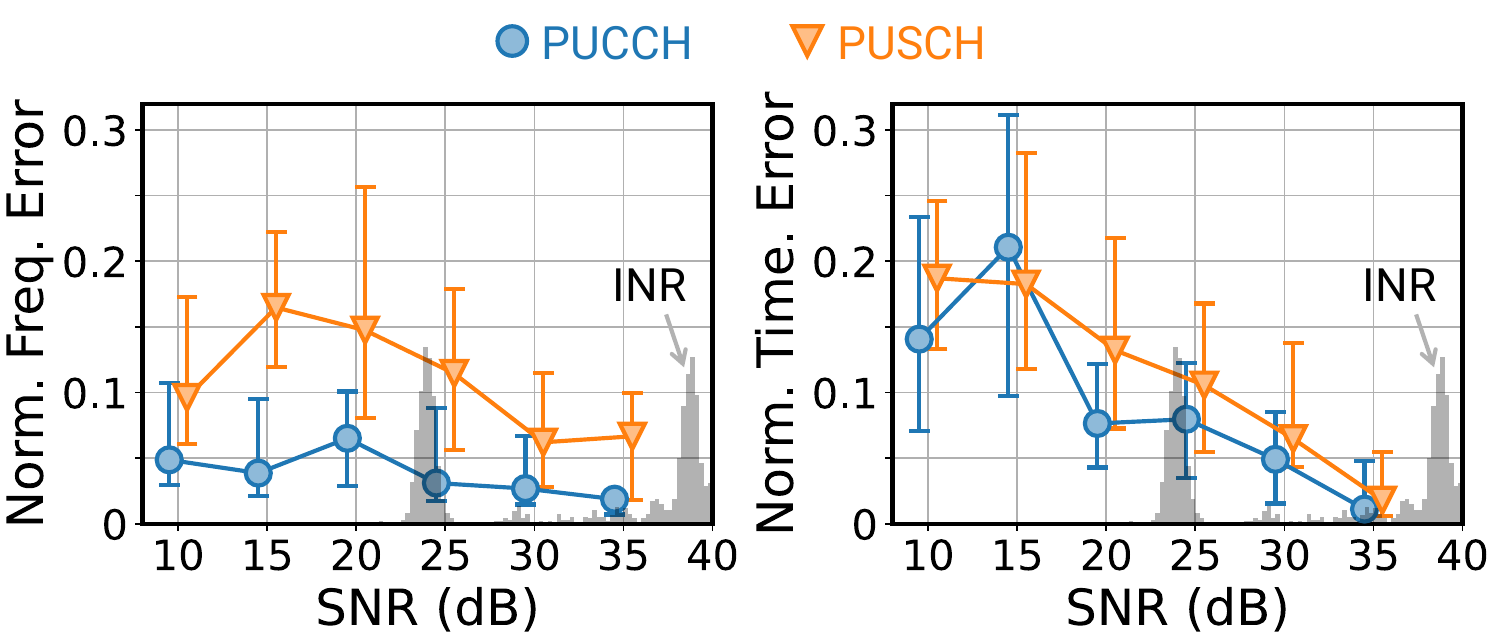}
    \vspace{-3mm}
    \caption{The normalized frequency/time median and 25/75-th quantile errors averaged over radar types by {\namebf}.}
    \vspace{-5mm}
    \label{fig:line-snr-freq-time}
\end{figure}

\myparatight{Frequency/Time Localization errors.}
We further average the median frequency/time localization errors over the five radar types and plot their tendencies over different SNRs. Overall, when decreasing the SNRs, both frequency and time localization errors increase; under the same SNR, the frequency and time localization errors are lower on the PUCCH dataset than those on the PUSCH dataset.
As shown in Fig.~\ref{fig:line-snr-freq-time}(\emph{left}), the normalized frequency errors increase to {4.87/9.62\%} under {10}\thinspace{dB} SNR for the PUCCH and PUSCH datasets, respectively.
As for the starting time localization in Fig.~\ref{fig:line-snr-freq-time}(\emph{right}), the normalized time errors drop to {14.07/18.71\%} under {10}\thinspace{dB} SNR for the PUCCH and PUSCH datasets.

\subsection{System Model Efficiency}

\begin{figure}[!t]
    \centering
    \includegraphics[width=0.95\columnwidth]{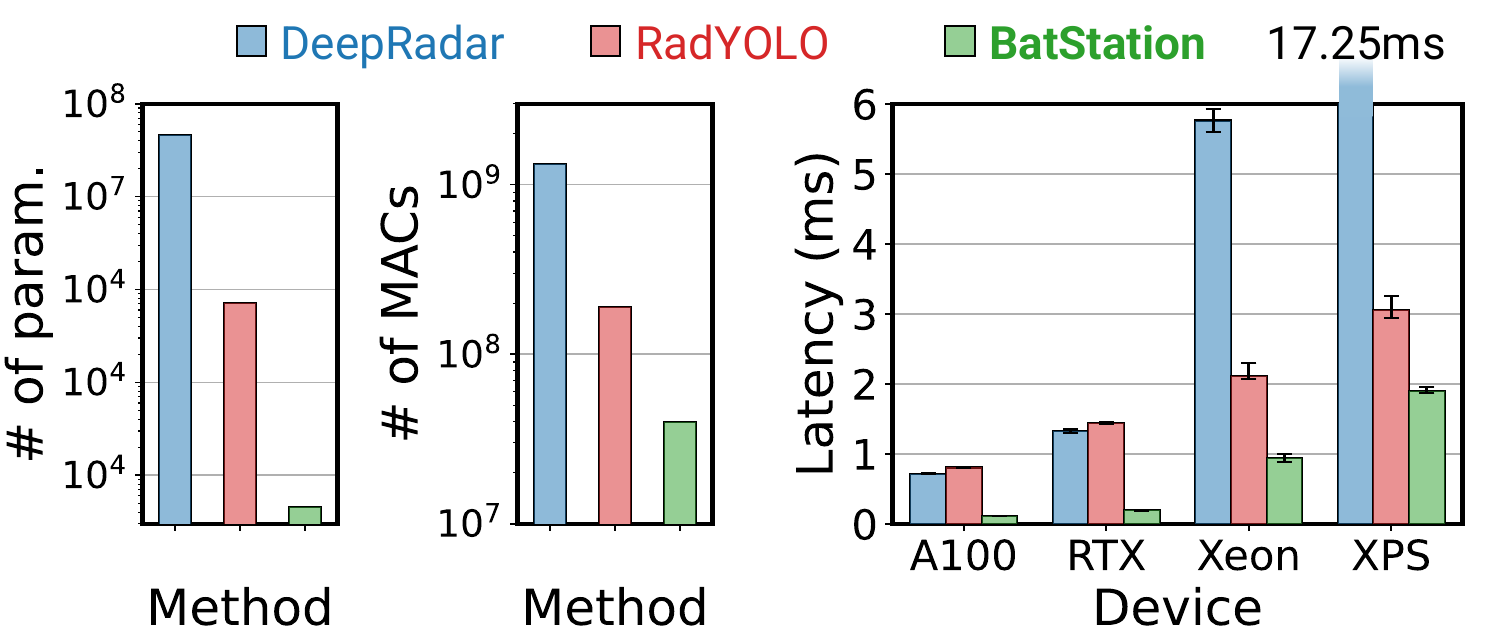}
    \vspace{-3mm}
    \caption{The model size, number of MACs per inference, comparison of {\namebf} and the two ML-driven baselines, and their runtime latency on various computing platforms.}
    \vspace{-5mm}
    \label{fig:latency-test}
\end{figure}

\myparatight{Model size.}
We show the number of trainable parameters in Fig.~\ref{fig:latency-test}(\emph{left}), where {\name} only has {4,560} trainable parameters across all trained templates, which is {0.637\%} of that in RadYOLO. This indicates {\name}'s robustness against the over-fitting issue, i.e., less reliance on the training dataset while more portable over various hardware on BSs.

\myparatight{Number of MACs per inference.}
As shown in Fig.~\ref{fig:latency-test}(\emph{middle}), {\name} only contains {40M} MACs per inference to process one uplink slot.
In contrast, DeepRadar and RadYOLO require {191M/1,322M} MACs plus non-linear activation functions (e.g., ReLU, Sigmoid).
This metric reflects the energy efficiency during the model inference.

\myparatight{Latency measurements.}
We measure the runtime latency on various computing devices, including two GPUs (NVIDIA A100, NVIDIA TITAN RTX) and two CPUs (Intel Xeon 6384 and Dell XPS), whose median latency with 25/75-th percentiles are shown in Fig.~\ref{fig:latency-test}(\emph{right}).
On the two GPUs, {\name} consumes a latency of {0.11/0.20}\thinspace{ms} on A100/TITAN RTX to process one uplink slot, {7.50/7.38}$\times$ faster than RadYOLO. This latency gain comes from the large kernel sizes of {\name}'s template correlation, which is compatible with the powerful parallelism of GPUs.
In addition, the latency of {\name} is {0.94}\thinspace{ms} on Intel Xeon 6384, i.e., {2.23}$\times$ faster than that of RadYOLO. This latency meets the real-time requirement when there are up to five uplink slots per transmission periodicity ({5}\thinspace{ms}).
On the Dell XPS, the latency is {1.90}\thinspace{ms} ({1.61}$\times$ faster than RadYOLO), supporting up to two uplink slots.
Such a latency benefit results from not only {\name}'s smaller number of MACs, but also the parallel computing paradigm over multiple radar templates instead of multiple serial layers as of the ML models.
\section{Conclusion}
\label{sec:conclusion}

In this paper, we propose {\name}, an in-situ radar sensing system that can be seamlessly integrated into 5G BSs.
Unlike prior ML-based approaches that rely on large-scale training datasets and heavy inference costs, {\name} achieves three key radar sensing tasks--existence detection, type classification, and time/frequency localization--leveraging novel radar signal separation from uplink resource grids and zero-shot template generation.
Extensive over-the-air experiments on a 5G SDR platform with commodity 5G traffic reveal {\name}'s radar sensing capability in real-world scenarios.

\section*{Acknowledgments}
The work was supported in part by NSF grants CNS-2112562, CNS-2128638, AST-2232458, ECCS-2434131, and CNS-2443137, and NVIDIA Academic Grants.

\bibliographystyle{ACM-Reference-Format}
\bibliography{reference}

\end{document}